\pdfoutput=1

\documentclass[letterpaper,twocolumn,10pt]{article}
\usepackage{usenix,epsfig,endnotes}
\usepackage{amsfonts}
\usepackage[square, comma, numbers,sort&compress]{natbib}

\usepackage[left=1in,top=1in,right=1in,bottom=1in,columnsep=0.25in]{geometry}
\usepackage{multirow}
\usepackage[pdfstartview=FitH, bookmarksnumbered=true, bookmarksopen=true, colorlinks=true, citecolor=blue, colorlinks=blue, linkcolor= blue, urlcolor=blue, draft]{hyperref}
\usepackage{soul}
\usepackage{url}
\usepackage[hang,scriptsize,tight,nooneline]{subfigure}

\usepackage{xspace}
\usepackage{flushend} % Balance columns on last page
\usepackage{titling}
\posttitle{\par\end{center}\vspace{-1.5em}}
\newcommand{\cut}[1]{}
\newcommand{\fixme}[1]{}

\newcommand{\new}[1]{{\textcolor{black}{#1}}}
\newcommand{\newest}[1]{{\textcolor{black}{#1}}}
\newcommand{\newx}[1]{{\textcolor{black}{#1}}}
\newcommand{\newestx}[1]{{\textcolor{black}{#1}}}

\newcommand{\cutatlastminute}[1]{}

\newcommand{\paragraphb}[1]{\vspace{0.03in}\noindent{\bf #1}}
\newcommand{\paragraphe}[1]{\vspace{0.03in}\noindent{\emph{#1}}}

\newcommand{\sys}{Jellyfish\xspace}
\newcommand{\ankitb}{\textcolor{black}}
\newcommand{\ankitx}{\textcolor{black}}

\newcommand{\cycready}{\textcolor{black}}

\setstcolor{red}

\begin{document}
\baselineskip=12bp
%%%%%%%%%%%%%%%%%%%%%%%%%%%%%%%%%%%%%%%%%%%%%%%%%%%%%%%%%%%%%%%%%%%%%%%%%%%%%%%%

%\title{A Random Mess Network}
\title{Jellyfish: Networking Data Centers Randomly}
%\thanks{Part of this work was published at HotCloud'11~\cite{jellyfish}.}
%\numberofauthors{2}
\author{
Ankit Singla$^\dag$\thanks{A coin toss decided the order of the first two authors.}, Chi-Yao Hong$^\dag$$^*$, Lucian Popa$^\sharp$, P. Brighten Godfrey$^\dag$\\
$^\dag$ University of Illinois at Urbana--Champaign\\
$^\sharp$ HP Labs%University of California, Berkeley\\
}

\date{}
\maketitle

%\thispagestyle{empty}
%\vspace{-0.80in}
\vspace{-30pt}
\abstract{
Industry experience indicates that the ability to incrementally 
expand data centers is essential. However, existing high-bandwidth
network designs have rigid structure that interferes with incremental 
expansion.  We present \sys, a high-capacity network interconnect
which, by adopting a random graph topology, yields itself naturally 
to incremental expansion.  Somewhat surprisingly, \sys is more 
cost-efficient than a fat-tree, supporting as many as 
$25\%$ more servers at full capacity using the same equipment at the scale of a few thousand nodes, and this 
advantage improves with scale. \sys also allows great 
flexibility in building networks with different degrees of 
oversubscription. However, \sys's unstructured design brings new 
challenges in routing, physical layout, and wiring. \newest{We describe 
approaches to resolve these challenges, and our evaluation 
suggests that \sys could be deployed in today's data centers.}
}

\section{Introduction}
\label{sec:intro}

A well provisioned data center network is critical
to ensure that servers do not face bandwidth bottlenecks to
utilization; to help isolate services from each other; and
to gain more freedom in workload placement, rather than having
to tailor placement of workloads to where bandwidth is 
available~\cite{inmyway}. As a result, a significant
body of work has tackled the problem of building high capacity network 
interconnects~\cite{fattree, vl2, bcube, portland, dcell, 
cthrough, helios, proteus}.

One crucial problem that these designs encounter is 
incremental network expansion, \emph{i.e.}, adding servers
and network capacity incrementally to the data center.
Expansion may be necessitated by growth of the user base, which requires more servers,
or by the deployment of more bandwidth-hungry applications. Expansion
within a data center is possible through either
planned overprovisioning of space and power, or by upgrading old 
servers to a larger number of more powerful but energy-efficient new servers.  \newest{Planned expansion is a practical strategy to reduce up-front capital expenditure~\cite{incremental}.}

Industry experience indicates that incremental expansion is 
important. Consider the growth of Facebook's data center 
server population from roughly $30$,$000$ in Nov. $2009$ to 
$>$$60$,$000$ by June $2010$~\cite{fbservers}. While Facebook 
has added entirely new data center facilities, much of this 
growth involves incrementally expanding existing facilities by ``adding
capacity on a daily basis"~\cite{dailycap}. For instance, Facebook 
announced that it would double the size of its facility at Prineville, Oregon by early $2012$~\cite{fbsvrcase}.
A $2011$ survey~\cite{dcnsurvey} of $300$ enterprises that run data centers of 
a variety of sizes found that \newest{$84\%$ of firms would probably or definitely  expand their data centers in $2012$.}
Several industry products advertise incremental expandability of the server pool,
including SGI's IceCube (marketed as ``The Expandable Modular
Data Center''~\cite{icecubeair}; expands $4$ racks at a time)
and HP's EcoPod~\cite{hp-ecopod} (a ``pay-as-you-grow'' enabling 
technology~\cite{hp-ecopod1}). \cut{However, in both cases, no mention 
is made of how the network supports such expansion of the server pool.}

\emph{Do current high-bandwidth data center network proposals allow incremental 
growth?} Consider the fat-tree interconnect, as proposed in \cite{fattree}, as an illustrative example.
The entire structure is completely determined by the port-count of the
switches available. This is limiting in at least two ways. First, it
makes the design space very coarse: full bisection bandwidth fat-trees 
can only be built at sizes $3456$, $8192$, $27648$, and $65536$ corresponding
to the commonly available port counts of $24$, $32$, $48$, and $64$\footnote{Other topologies have similar problems: a hypercube~\cite{hypercubenetwork} 
allows only power-of-$2$ sizes, a de Bruijn-like construction~\cite{lucian10} 
allows only power-of-$3$ sizes, etc.}.
Second, even if (for example) $50$-port switches were available, the smallest ``incremental''
upgrade from the $48$-port switch fat-tree would add $3$,$602$ servers and would require replacing every switch. 

There are, of course, some workarounds. One can replace a switch with one of larger port count or oversubscribe certain switches, but this makes capacity distribution constrained and uneven across the servers. One could leave free ports for future network connections~\cite{dcell,legup} but this wastes investment until actual expansion. Thus, without compromises on bandwidth or cost, such topologies are not amenable to incremental growth.

% There is, of course, the possibility of making localized changes like replacing a switch with one with larger port count; however, this necessarily makes capacity distribution unfair across the server pool. The only prior work~\cite{legup} that directly addresses the problem of incremental expansion, attempts to make the most out of this bad situation -- it searches for optimum additions of network equipment to Clos networks. In contrast, we design for expansion, resulting (as we show in \S\ref{sec:flexibility}) in significant gains in network capacity for the same (expanding) data center under the same budgetary constraints.

\cut{
of making localized changes like replacing a switch with one with larger port
count -- this necessarily requires either the servers in the vicinity to
share capacity unevenly compared to the rest of the server pool, or 
greater expense to ensure that this capacity is at par with the rest of
the network. Thus, without compromise on its structure, bandwidth or cost, 
a fat-tree is not amenable to incremental
growth.}

%An alternative approach suggested in the literature~\cite{dcell}, is based on leaving free ports for future network connections. But the cost of these free ports, is an unnecessary sunk investment for the period in which the network does not expand. Thus, without compromises on bandwidth or cost, such topologies are not amenable to incremental growth.

Since it seems that \emph{structure} hinders incremental expansion, we
propose the opposite: a random network interconnect.
The proposed interconnect, which we call {\bf \sys}, is a \emph{degree-bounded\footnote{\newest{Degree-bounded means that the number 
of connections per node is limited, in this case by switch port-counts.}} random graph} 
topology among top-of-rack (ToR) switches.  The inherently sloppy nature
of this design has the potential to be significantly more flexible than 
past designs.  Additional components---racks of servers or switches to improve 
capacity---can be incorporated with a few random edge swaps.  The 
design naturally supports heterogeneity, allowing the addition of newer
network elements with higher port-counts as they become available, unlike past 
proposals which depend on certain regular port-counts~\cite{fattree, vl2, 
portland, dcell, bcube, proteus}. \sys also allows construction of arbitrary-size
networks, unlike topologies discussed above which limit the network to
very coarse design points dictated by their structure.

\new{Somewhat surprisingly, \sys supports {\em more} servers  than a fat-tree~\cite{fattree} built using the same network equipment
while providing at least as high per-server bandwidth, measured either via bisection bandwidth or in throughput under a random-permutation traffic pattern.} In addition, \sys has lower mean path length, and is resilient to failures and miswirings.

But a data center network that lacks regular structure is a somewhat radical departure from traditional designs, and this presents several important challenges that must be addressed for \sys to be viable. Among these are routing (schemes depending on a structured topology are not applicable), physical construction, and cabling layout. We describe simple approaches to these problems which suggest that \sys could be effectively deployed in today's data centers.

Our key contributions and conclusions are as follows:

\begin{itemize}
	\item We propose \sys, an incrementally-expandable, high-bandwidth data center interconnect based on a random graph.
	
	\item \new{We show that \sys provides quantitatively easier incremental expansion than prior work on incremental expansion in Clos networks~\cite{legup}, growing incrementally \cut{to a slightly higher capacity network} at only $40\%$ of the expense of~\cite{legup}.}

	\item We conduct a comparative study of the bandwidth of several proposed data center network topologies.  Jellyfish can support $25$\% more servers than a fat-tree while using the same switch equipment and providing at least as high \new{bandwidth.  This advantage increases with network size and switch port-count.}  Moreover, we propose {\em degree-diameter optimal graphs}~\cite{degree-diameter} as \newest{benchmark topologies for high capacity at low cost}, and show that \sys remains within $10$\% of these carefully-optimized networks.

	%\item Motivated by the question of how close \sys's bandwidth is to optimal, we suggest {\em degree-diameter optimal graphs} as candidate benchmark topologies.  We show these graphs indeed perform even better than \sys --- but \sys remains within $5$-$10$\% of these carefully-optimized topologies.
	
	\item \new{Despite its lack of regular structure, packet-level simulations show that \sys's bandwidth can be effectively utilized via existing forwarding technologies that provide high path diversity.}
	
	%\item We demonstrate in packet-level simulations that \sys's bandwidth can be effectively utilized via a practical (indeed, already implemented!) technique, multipath TCP~\cite{mptcp} --- despite the lack of regular structure that is sometimes used to ease routing in other topologies.

	\item We discuss effective techniques to realize physical layout and cabling of \sys. \sys may have higher cabling cost \newx{than other topologies, since its cables can be longer}; but when we restrict \sys to use cables of length similar to the fat-tree, it still improves on the fat-tree's \new{throughput}.

\end{itemize}

\paragraph{Outline:} Next, we discuss related work (\S\ref{sec:related}),
followed by a description of the \sys topology (\S\ref{sec:topology}), and an 
evaluation of the topology's properties, unhindered by routing and congestion
control (\S\ref{sec:researchopps}). We then evaluate the topology's performance
with routing and congestion control mechanisms (\S\ref{sec:routing}). We discuss effective cabling schemes and 
physical construction of \sys in various deployment scenarios (\S\ref{sec:cabling}),
and conclude (\S\ref{sec:conclusion}).

\section{Related Work}
\label{sec:related}

Several recent proposals for high-capacity networks 
exploit special structure for topology and routing. These include folded-Clos (or fat-tree)
designs~\cite{fattree, portland, vl2}, several designs that
use servers for forwarding~\cite{bcube, dcell, mdcube},
and designs using optical networking technology~\cite{helios,cthrough}.
High performance computing literature has also studied 
carefully-structured expander graphs~\cite{leighton}.

However, none of these architectures address the
\emph{incremental expansion} problem. For some
(including the fat-tree), adding servers while 
preserving the structural properties would require
replacing a large number of network elements and extensive rewiring. 
MDCube~\cite{mdcube} allows expansion at
a very coarse rate (several thousand servers). DCell and 
BCube~\cite{dcell,bcube} allow expansion to an \emph{a priori} known target 
size, but require servers with free ports 
reserved for planned future expansion.

Two recent proposals, Scafida~\cite{scafida} (based on scale-free graphs)
and Small-World Datacenters (SWDC)~\cite{swdc}, are similar to \sys in that they employ randomness, but are significantly different than our design because they require correlation 
(i.e., structure) among edges. This structured design makes it unclear whether the topology
retains its characteristics upon incremental expansion; neither proposal investigates
this issue.
 Further, in SWDC, the use of a regular lattice underlying
the topology creates familiar problems with incremental expansion.\footnote{For 
instance, using a $2$D torus as the lattice implies that maintaining the network
structure when expanding an $n$ node network requires adding 
$\Theta(\sqrt n)$ new nodes. The higher the dimensionality of the lattice, the 
more complicated expansion becomes.} \sys also has a capacity advantage
over both proposals: Scafida has marginally worse bisection bandwidth
and diameter than a fat-tree, while \sys improves on fat-trees on 
both metrics. We show in \S\ref{sec:efficiency} that \sys 
has higher bandwidth than SWDC topologies built using the 
same equipment.

LEGUP~\cite{legup} attacks the expansion problem
by trying to find optimal upgrades for Clos networks. 
However, such an approach is fundamentally limited by having 
to start from a rigid structure, and adhering to it during 
the upgrade process. Unless free ports are preserved for
such expansion (which is part of LEGUP's approach), this can
cause significant overhauls of the topology even when adding 
just a few new servers. In this paper, we show that \sys 
provides a simple method to expand the network to almost any
desirable scale. \cut{In addition, we show that \sys networks are 
inherently less expensive than fat trees, which are representative
Clos topologies (widely used for datacenter interconnects~\cite{fattree, vl2, portland}).}
Further, our comparison with LEGUP (\S\ref{sec:flexibility}) over 
a sequence of network expansions illustrates that \sys provides 
significant cost-efficiency gains in incremental expansion.

REWIRE~\cite{rewire} is a heuristic optimization method to 
find high capacity topologies with a given cost budget, taking into account length-varying cable cost.  While~\cite{rewire} compares with random graphs, the results are inconclusive.\footnote{REWIRE attempts to improve a given ``seed'' graph. The seed could be a random graph, so in principle~\cite{rewire} should be able to obtain results at least as good as \sys.  In~\cite{rewire} the seed was an empty graph.  The results show, in some cases, fat-trees obtaining more than an order of magnitude worse bisection bandwidth than random graphs, which in turn are more than an order of magnitude worse than REWIRE topologies, all at equal cost.  In other cases, \cite{rewire} shows random graphs that are disconnected.  These significant discrepancies could arise from (a) separating network port costs from cable costs rather than optimizing over the total budget, causing the random graph to pay for more ports than it can afford cables to connect; (b) assuming linear physical placement of all racks, so cable costs for distant servers scale as $\Theta(n)$ rather than $\Theta(\sqrt n)$ in a more typical two-dimensional layout; and (c) evaluating very low bisection bandwidths (\textbf{0.04} to $0.37$) --- in fact, at the highest bisection bandwidth evaluated, \cite{rewire} indicates the random graph has higher throughput than REWIRE. The authors indicated to us that REWIRE has difficulty scaling beyond a few hundred nodes.} Due to the recency of~\cite{rewire}, we have left a direct quantitative comparison to future work.

Random graphs have been examined in the context of 
communication networks~\cite{rrgnetwork} previously.  Our contribution lies in applying random graphs to allow incremental 
expansion and in quantifying the efficiency 
gains such graphs bring over traditional data center topologies.

\section{\sys Topology}
\label{sec:topology}

\paragraphb{Construction:}
The \sys approach is to construct a random graph at the
top-of-rack (ToR) switch layer. Each ToR switch $i$ has some number $k_i$
of ports, of which it uses $r_i$ to connect to other ToR switches, and
uses the remaining $k_i-r_i$ ports for servers. In the simplest case,
which we consider by default throughout this paper, every switch has the
same number of ports and servers: $\forall i$, $k=k_i$ and $r=r_i$. With
$N$ racks, the network supports $N (k - r)$
servers. In this case, the network is a {\em random regular graph}, which
we denote as RRG($N$, $k$, $r$).  This is a well known construct in graph theory and has
several desirable properties as we shall discuss later.

Formally, RRGs are sampled uniformly from the space of all $r$-regular graphs. This is a complex problem in graph theory \cite{rrggen}; however, a simple procedure produces ``sufficiently uniform'' random graphs which empirically have the desired performance properties. One can simply pick a random pair of switches with free 
ports \cycready{(for the switch-pairs are not already neighbors)}, join them with a link, and repeat until no further links can be added. If a switch remains with $\geq2$ free ports $(p_1, p_2)$ --- which includes the case of incremental expansion by adding a new switch --- these can be incorporated by removing a uniform-random existing link $(x,y)$, and adding links $(p_1, x)$ and $(p_2,y)$.  Thus only a single unmatched port might remain across the whole network.

Using the above idea, we generate a blueprint
for the physical interconnection. (Allowing human 
operators to ``wire at will'' may result in poor topologies due to 
human biases -- for instance, favoring shorter cables
over longer ones.) We discuss cabling later in \S\ref{sec:cabling}.

\begin{figure*}
%\vspace{-12pt}
\centering
\subfigure[]{ \label{fig:fat_viz} \includegraphics[width=1.8in]{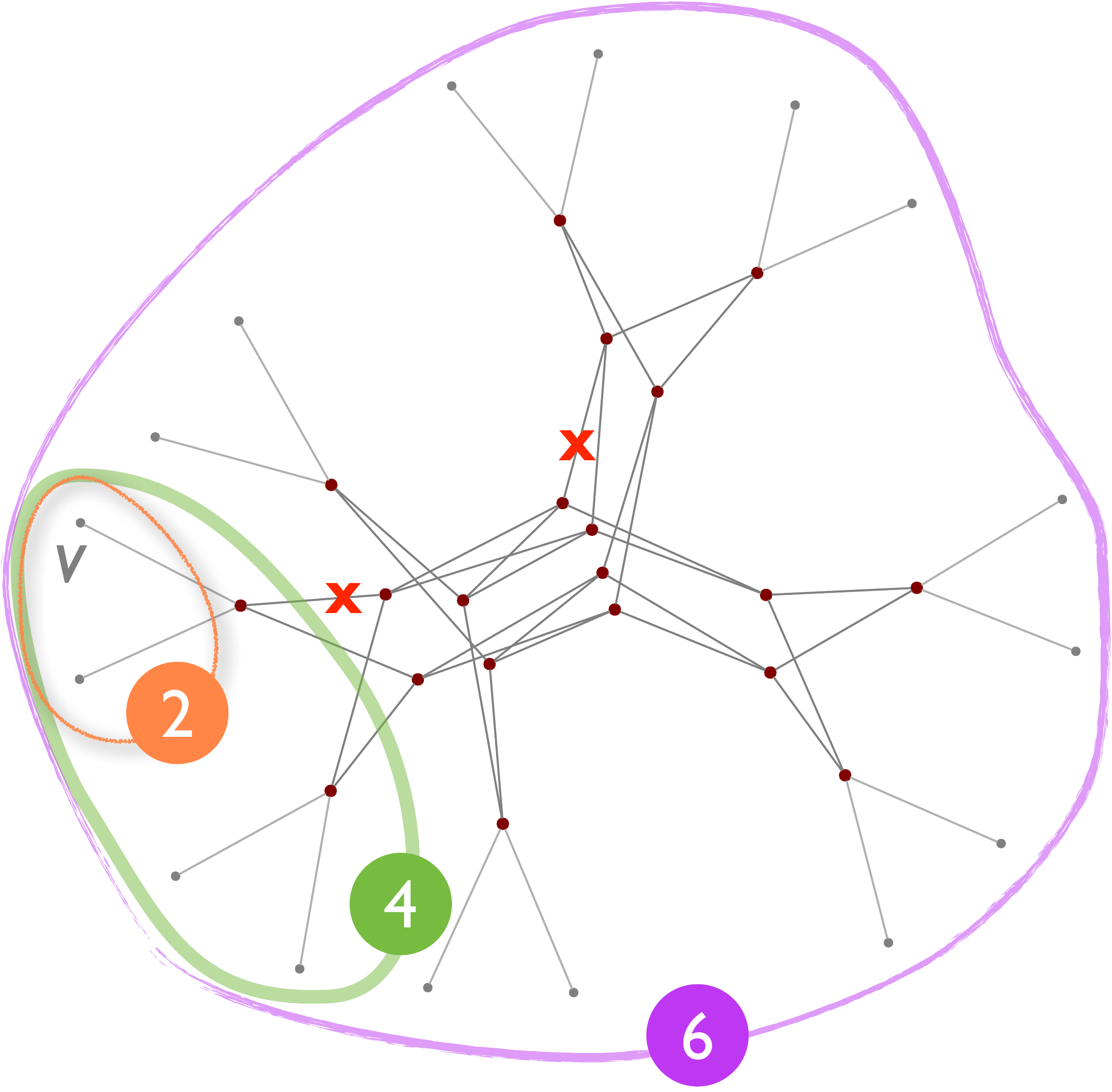}}
\hspace{10pt}
\subfigure[]{ \label{fig:jelly_viz} \includegraphics[width=1.8in]{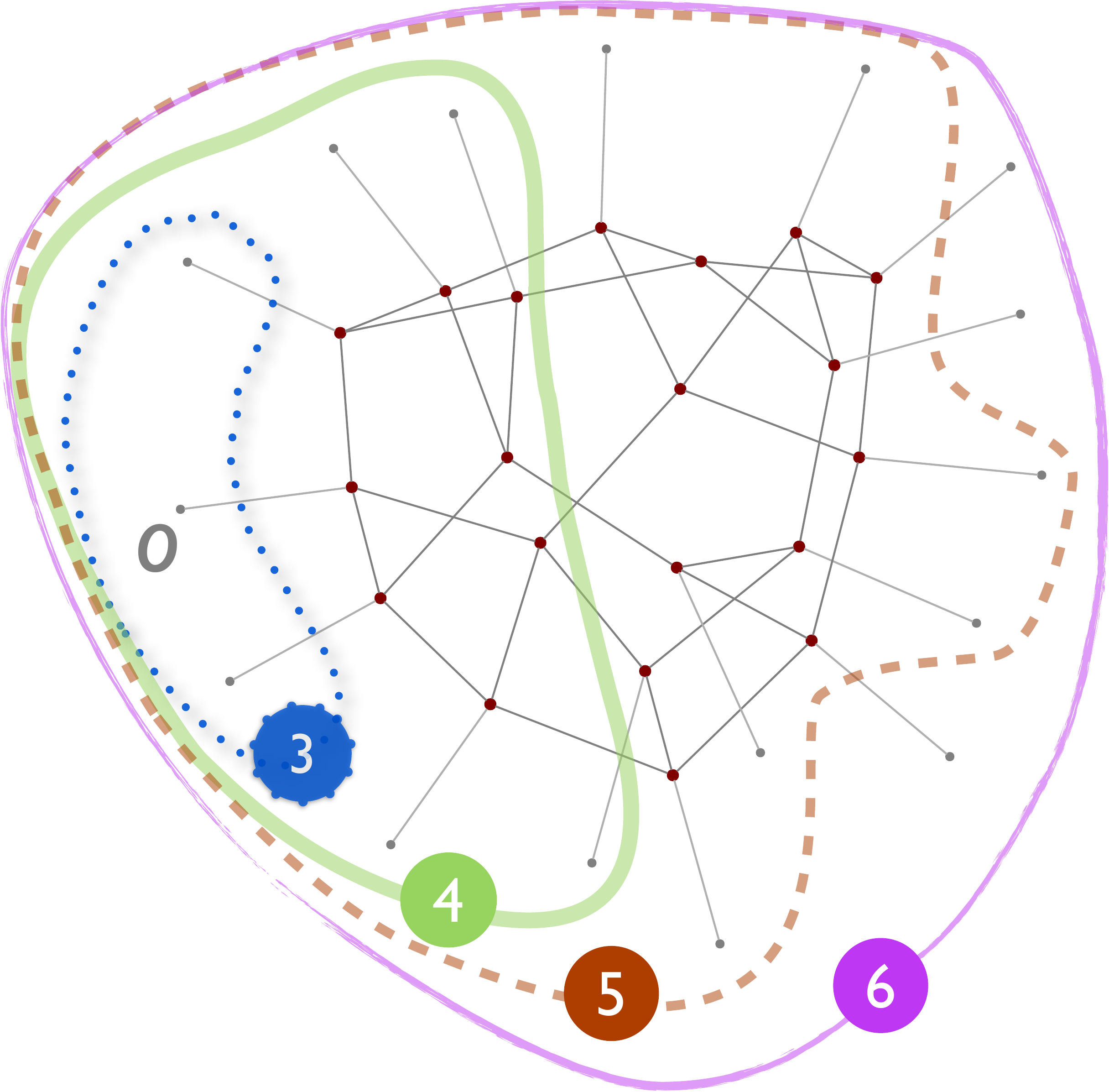}}
\hspace{6pt}
\subfigure[]{ \label{fig:cdf_pathlengthe} \includegraphics[width=2.43in]{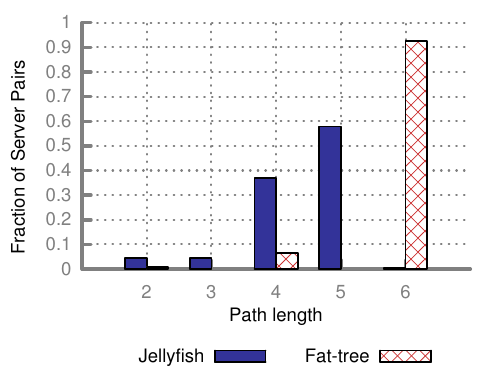}}
\caption{\small \em Random graphs have high throughput because they have low average path length, and therefore do less work to deliver each packet. \textbf{(a):} Fat-tree with $16$ servers and $20$ four-port switches. \textbf{(b):} \sys  with identical equipment. The servers are leaf nodes; switches are interior nodes. \newest{Each `concentric' ring contains servers reachable from any server $v$ in the fat-tree, and an arbitrary server $o$ in \sys, within the number of hops in the marked label.} \sys can reach many more servers in few hops because in the fat tree, many edges (like those marked ``X'') are redundant from a path-length perspective. \textbf{(c):} Path length distribution between servers for a $686$-server \sys (drawn from $10$ trials) and same-equipment fat-tree.}
\vspace{-12pt}
\end{figure*}

\paragraphb{Intuition:} Our two key goals are high bandwidth and flexibility.  The intuition for the latter property is simple: lacking structure, the RRG's network capacity becomes ``fluid'', easily wiring up any number of switches, heterogeneous degree distributions, and newly added switches with a few random link swaps.

But why should random graphs have high bandwidth? We show quantitative results later, but here we present the intuition. \newestx{The end-to-end throughput of a topology depends not only on the capacity of the network, but is also inversely proportional} to the {\em amount of network capacity consumed to deliver each byte} --- that is, the \newestx{average} path length.  Therefore, assuming that the routing protocol is able to utilize the network's full capacity, low \newestx{average} path length allows us to support more flows at high throughput.  To see why \sys has low path length, Fig.~\ref{fig:fat_viz} and~\ref{fig:jelly_viz} visualize a fat-tree and a representative \sys topology, respectively, with \newestx{\emph{identical}} equipment. \newestx{Both topologies have diameter $6$, meaning that any server can reach all other servers in $6$ hops.} \newestx{However, in the fat-tree, each server can only reach $3$ others in $\leq 5$ hops.} In contrast, in the random graph, the typical origin server labeled $o$ can reach $12$ servers in $\leq 5$ hops, and $6$ servers in $\leq 4$ hops. The reason for this is that many edges in the fat-tree are not useful from the perspective of their effect on path length; for example, deleting the two edges marked X in Fig.~\ref{fig:fat_viz} does not increase the path length between any pair of servers. In contrast, the RRG's diverse random connections lead to lower mean path length.\footnote{This is related to the fact that RRGs are \emph{expander graphs}~\cite{rrgexpander}.}
\cut{Intuitively, a small average path length implies that each flow takes a small number of hops, thus consuming a small amount of network capacity and allowing the network to support a large number of flows at high throughput.} \newest{Figure~\ref{fig:cdf_pathlengthe} demonstrates these effects at larger scale. With $686$ servers, $>$$99.5\%$ of source-destination pairs in \sys can be reached in fewer than $6$ hops, while the corresponding number is only $7.5\%$ in the fat-tree.}

\cut{
\emph{Any} topology with a small 
average path length and a sufficient number of edges should provide 
high bandwidth connectivity. !!! Not quite true -- there might be
a bottleneck.  Imagine two cliques connected by a single edge.  Pretty
low path length, lots of edges...horrible bisection bandwidth.

This follows from the 
observation that short paths consume resources on
a small number of network edges. Robust theoretical proofs for
almost every RRG to have low diameter and high bisection bandwidth
are well known~\cite{bwbound, diabound}. 
}

\cut{Our construction maintains a list of nodes together with the number
of remaining connections for each node. This list is initialized with all
$N$ nodes, with each remaining-connection-number set to the number of
network-ports $r$. A pair of nodes is picked uniform-randomly from the
list and connected if there isn't already have an edge between them. The
remaining connections for both these nodes are reduced, and nodes with
$0$ remaining connections are removed from the list. This process
continues until no further connections can be made.}

%i.e., a (possibly $0$-size) clique of nodes is left in the list. 
%Degenerate cases, like a large number of nodes left in the list at
%termination, are low-probability events. In these cases, 
%this construction algorithm can be rerun from scratch. A small number of nodes with a few connections remaining do not

\cut{While it is true that
our construction may not yield truly uniform-random regular graphs,}

\section{\sys Topology Properties}
\label{sec:researchopps}
\vspace{-7pt}

This section evaluates the efficiency, flexibility and resilience of
\sys and other topologies. Our goal is to measure the raw capabilities of the topologies,
were they to be coupled with optimal routing and congestion control.
We study how to perform routing and congestion control separately, in \S\ref{sec:routing}.

\vspace{8pt}

\noindent Our key findings from these experiments are:

\begin{itemize}\addtolength{\itemsep}{-0.5\baselineskip}
\item \sys can support $27\%$ more servers at full capacity than a
(same-switching-equipment) fat-tree at a scale of $<$$900$ servers. 
The trend is for this advantage to \emph{improve} with scale.
\item \sys's network capacity is $>$$91\%$ of the best-known
degree-diameter graphs\newx{~\cite{degree-diameter}, which we propose as benchmark bandwidth-efficient graphs}.
\item Paths are shorter on average in \sys than in a fat-tree, and the
{\em maximum} shortest path length (diameter) is the same or lower for all scales we tested.
\item Incremental expansion of \sys produces topologies identical in
throughput and path length \sys topologies
generated from scratch.
\item \sys provides a significant cost-efficiency advantage
over prior work (LEGUP~\cite{legup}) on incremental network expansion in Clos networks. In a
network expansion scenario that was made available for us to test, 
\sys builds a slightly higher-capacity expanded network at only $40\%$ of 
LEGUP's expense.
\item \sys is highly failure resilient, even more so than the
fat-tree. Failing a random $15\%$ of all links results in a
capacity decrease of $<$$16\%$.
\end{itemize}

\paragraphb{Evaluation methodology:} Some of the results for network
capacity in this section\cut{(Fig.~\ref{fig:bwall},~\ref{fig:bwcomp})} are
based on explicit calculations of the theoretical bounds for bisection
bandwidth for regular random graphs.

All throughput results presented in this section are based on calculations of throughput for a specific class of traffic demand matrices with optimal routing.  The traffic matrices we use are {\em random permutation traffic}: each server sends at its full output link rate to a single other server, and receives from a single other server, and this permutation is chosen uniform-randomly.  Intuitively, random permutation traffic represents the case of no locality in traffic, as might arise if VMs are placed without regard to what is convenient for the network\footnote{Supporting such flexible network-oblivious VM placement without a performance penalty is highly desirable~\cite{inmyway}.}.
Nevertheless, evaluating other traffic patterns is an important question that we leave for future work. \cut{We also note that in a study of several traffic patterns in fat-trees~\cite{fattree}, random permutation was of intermediate difficulty among the patterns evaluated, with several patterns producing lower or higher throughput.}

Given a traffic matrix, we characterize a topology's raw capacity with \newest{``ideal'' load balancing by treating flows as splittable and fluid}. This corresponds to solving a standard multi-commodity network flow problem. (We use the CPLEX linear program solver~\cite{cplex}.)

For all throughput comparisons, we use the \emph{same switching equipment} 
(in terms of both number of switches, and ports on each switch) for each set of 
topologies compared. Throughput results are always normalized to $[0,1]$, and averaged over all flows.

For comparisons with the full bisection bandwidth fat-tree, we attempt to find, using a binary search procedure, the 
maximum number of servers \sys can support using the same switching equipment
as the fat-tree while satisfying the full traffic demands.  Specifically, 
each step of the binary search checks a certain number of servers $m$ by 
sampling three random permutation traffic matrices, and checking whether 
\sys supports full capacity for {\em all} flows in {\em all} three matrices.  
If so, we say that \sys supports $m$ servers at full capacity. After our 
binary search terminates, we verify that the returned number of servers is able to get full
capacity over each of $10$ more samples of random permutation traffic matrices.

\subsection{Efficiency}
\label{sec:efficiency}
%\vspace{-7pt}

\paragraphb{Bisection Bandwidth vs. Fat-Tree:} Bisection bandwidth, a common measure of network capacity, is the \emph{worst-case} bandwidth spanning any two equal-size partitions of a network.  Here, we compute the fat-tree's bisection bandwidth directly from its parameters; for \sys, we model the network as a RRG and apply a lower bound of Bollob\'{a}s~\cite{bwbound}.  We normalize bisection bandwidth by dividing it by the total line-rate bandwidth of the servers in one partition\footnote{Values larger than $1$ indicate overprovisioning.}.

\begin{figure}[!t]
\centering
\subfigure[]{ \label{fig:bwall}\includegraphics[width=3.1in]{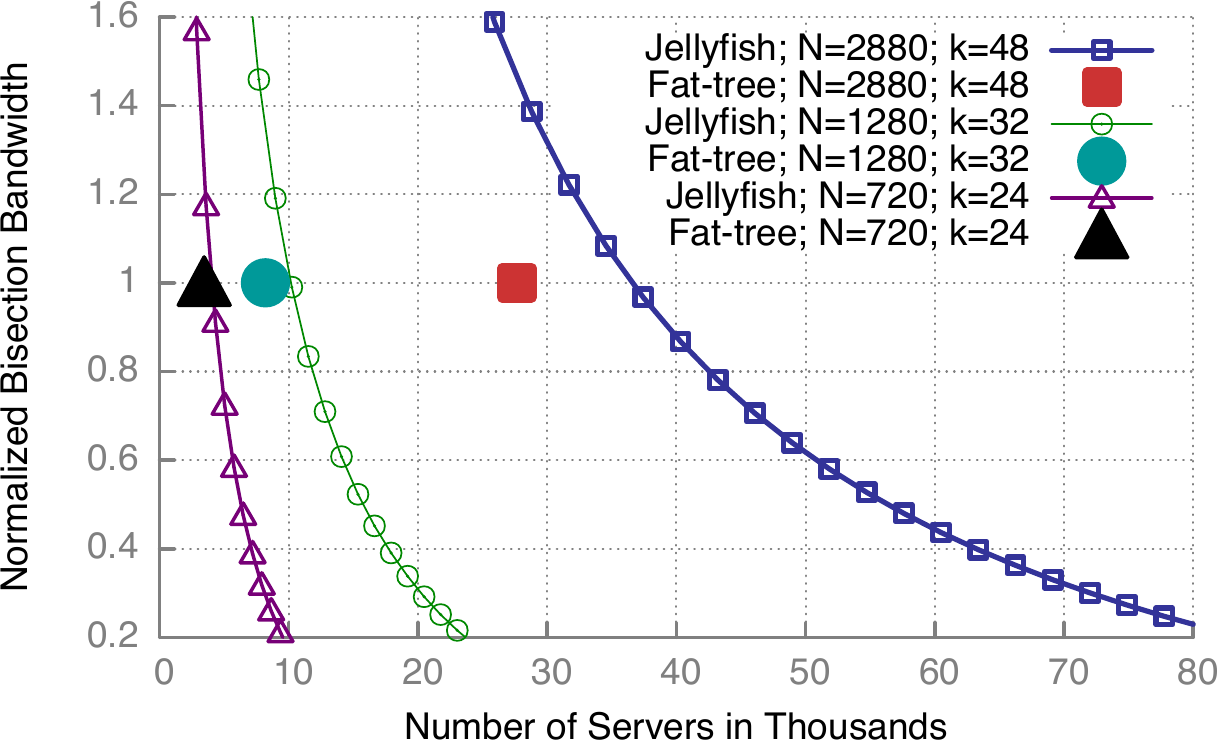}}
\vspace{-8pt}
\subfigure[]{ \label{fig:bwcomp}\includegraphics[width=3.1in]{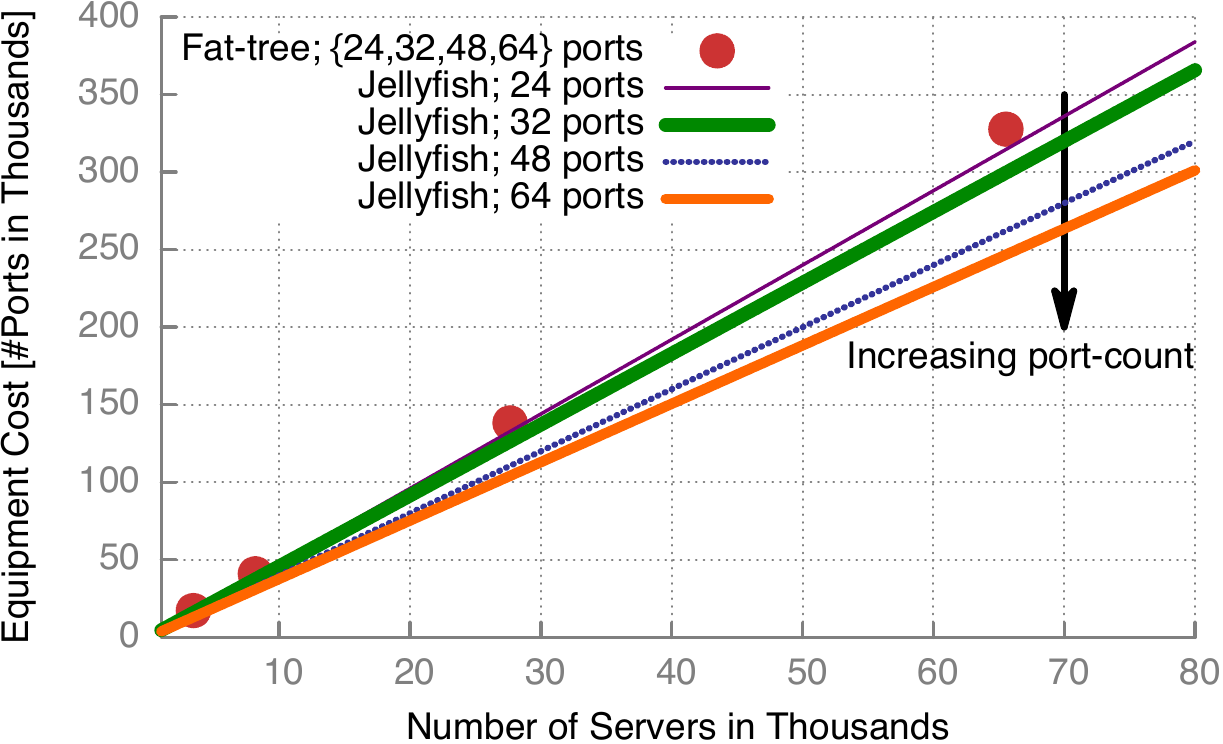}}
\vspace{-8pt}
\subfigure[]{ \label{fig:fat_cplex}\includegraphics[width=3.1in]{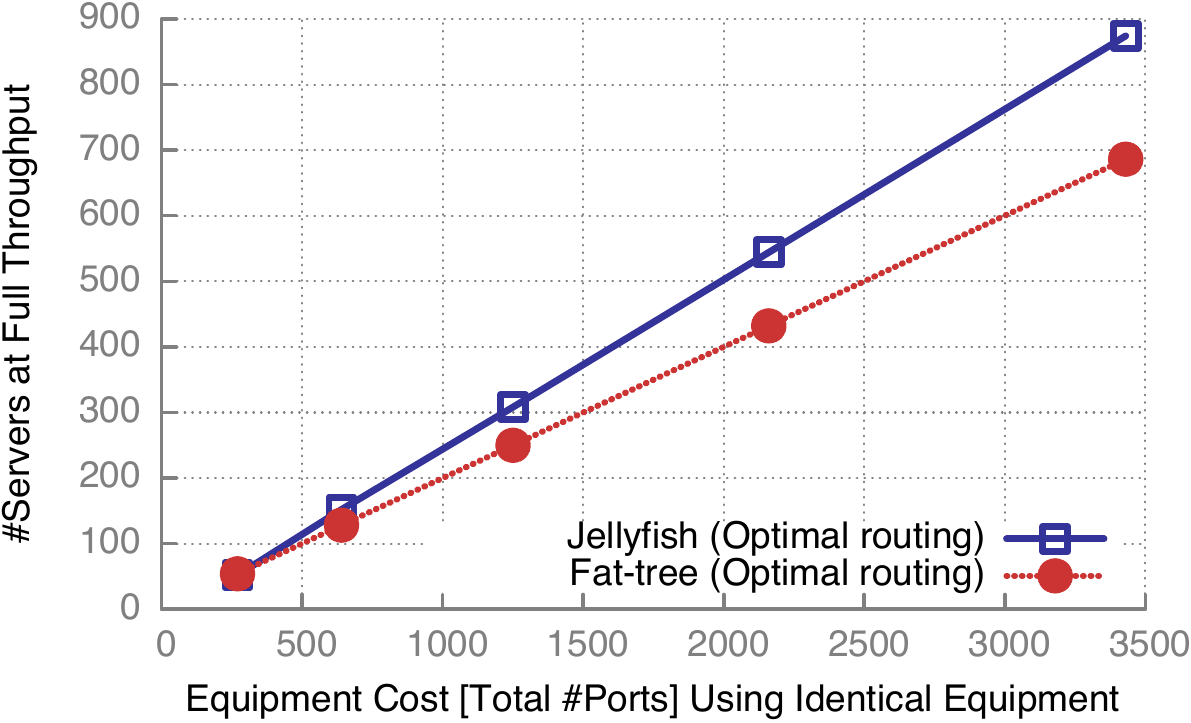}}
\caption{\small \em \sys offers virtually continuous design space, and packs in
more servers at high network capacity at the same expense as a fat-tree.
From theoretical bounds: (a) Normalized bisection bandwidth versus the 
number of servers supported; equal-cost curves, and (b) Equipment cost
versus the number of servers for commodity-switch port-counts ($24$, $32$, $48$\cut{, $64$, $96$, $128$}) at full bisection bandwidth. Under optimal routing, with
random-permutation traffic: (c) Servers supported at full capacity 
with the same switching equipment, for $6$, $8$, $10$, $12$ and $14$-port switches. 
Results for (c) are averaged over $8$ runs.}
\label{fig:bisectionBound}
\vspace{-15pt}
\end{figure}

Fig.~\ref{fig:bwall} shows that at the same cost, \sys supports a larger number of servers ($x$ axis) at full bisection bandwidth ($y$ axis $=1$). For instance, at the same cost as a fat-tree with $16$,$000$ servers, \sys can support $>$$20$,$000$ servers at full bisection bandwidth. Also, \sys allows the freedom to accept lower bisection bandwidth in exchange for supporting more servers or cutting costs by using fewer switches.

Fig.~\ref{fig:bwcomp} shows that the cost of building a full bisection-bandwidth 
network increases more slowly with the number of servers for \sys than for the 
fat-tree, especially for high port-counts. Also, the design choices for \sys 
are essentially continuous, while the fat-tree (following the design of~\cite{fattree}) allows only certain discrete 
jumps in size which are further restricted by the port-counts of available 
switches. (Note that this observation would hold even for over-subscribed
fat-trees.)

\sys's advantage increases with port-count, approaching twice the fat-tree's bisection bandwidth. To see this, note that the fat-tree built using $k$-port switches has $k^3/4$ servers, and being a full-bisection interconnect, it has $k^3/8$ edges crossing each bisection. \cut{Since it has $N=5k^2/4$ switches and thus} \newest{The fat-tree has $k^3/2$ switch-switch links, implying that its bisection bandwidth represents $\frac{1}{4}$ of its switch-switch links. For \sys, in expectation, $\frac{1}{2}$ of its switch-switch links cross any given bisection of the switches, which is \emph{twice} that of the fat-tree assuming they are built with the same number of switches and servers.} Intuitively, \sys's worst-case bisection should be slightly worse than this average bisection. The bound of \cite{bwbound} bears this out: in almost every $r$-regular graph with $N$ nodes, every set of $N/2$ nodes is joined by at least $N (\frac{r}{4} - \frac{\sqrt{r \ln 2}}{2})$ edges to the rest of the graph. As the number of network ports $r\to\infty$ this quantity approaches $Nr/4$, i.e., $\frac{1}{2}$ of the $Nr/2$ links.

\paragraphb{Throughput vs. Fat Tree:}
Fig.~\ref{fig:fat_cplex} uses the random-permutation traffic model to find the
number of servers \sys can support at full capacity, matching the fat-tree in capacity and switching equipment. The improvement is as much as $27\%$ more servers than the
fat-tree at the largest size ($874$ servers) we can use CPLEX to evaluate. 
\newx{As with results from Bollob\'{a}s' theoretical lower bounds on bisection bandwidth 
(Fig.~\ref{fig:bwall}, \ref{fig:bwcomp}),} the trend indicates that this improvement increases
with scale.

\paragraphb{Throughput vs. Degree-Diameter Graphs:} We compare \sys's capacity with that of the best known degree-diameter graphs. Below, we briefly explain what these graphs are, and why this comparison makes sense.

There is a fundamental trade-off between the degree and diameter of a 
graph of a fixed vertex-set (say of size $N$). At one extreme 
is a clique --- maximum possible degree ($N - 1$), and minimum possible diameter ($1$).
At the other extreme is a disconnected graph with degree $0$ and diameter 
$\infty$. The problem of constructing a graph with maximum possible number
$N$ of nodes while preserving given diameter and degree bounds is known as the {\em degree-diameter problem} and has 
received significant attention in graph theory. The problem is quite
difficult and the optimal graphs are only known for very small sizes: the largest degree-diameter
graph known to be optimal has $N=50$ nodes, with degree $7$ and diameter $2$~\cite{degree-diameter}.  A collection of optimal and best known graphs for other degree-diameter combinations is maintained at~\cite{degree-diameter}.

The degree-diameter problem relates to our objective in that short average 
path lengths imply low resource usage and thus high network capacity.
Intuitively, the best known degree-diameter topologies should support a
large number of servers with high network bandwidth 
and low cost (small degree). While we note the distinction 
between average path length (which relates more closely to the 
network capacity) and diameter, degree-diameter
graphs will have small average path lengths too.

Thus, we propose the best-known degree-diameter graphs as a benchmark for
comparison. Note that such graphs do not meet our incremental expansion
objectives; we merely use them as a capacity benchmark for \sys topologies.
But these graphs (and our measurements of them) may be of independent interest since they could be deployed as highly efficient topologies in a setting where incremental upgrades are unnecessary, such as a pre-fab container-based data center.

\begin{figure}
\centering
\includegraphics[width=3.125in]{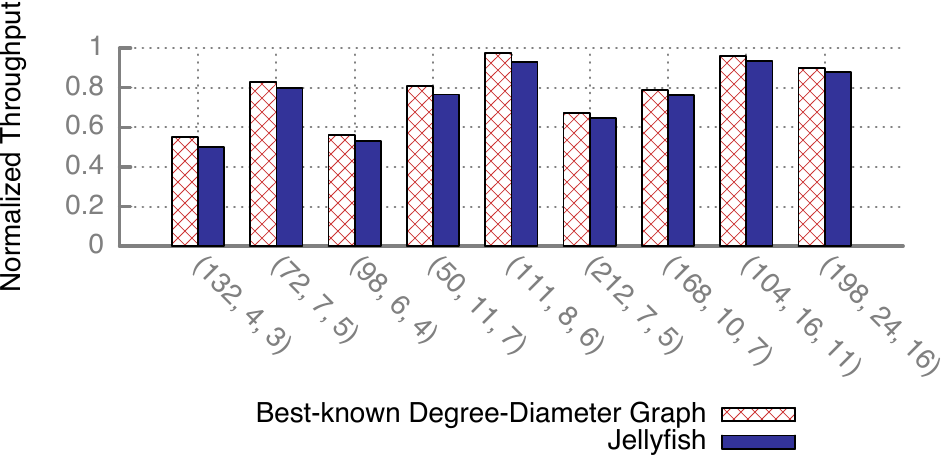}
\caption{\small \em \sys's network capacity is close to (i.e., $\sim$$91\%$ or more
in each case) that of the best-known degree-diameter graphs. The x-axis 
label ($A$, $B$, $C$) represents the number of switches ($A$), the switch 
port-count ($B$), and the network degree ($C$). Throughput is normalized against
the non-blocking throughput. Results are averaged over $10$ runs.\cut{For y-axis, we normalized the throughput 
by the full bisection rate. Each bar compares the normalized throughput obtained by \sys
to that obtained by using the best-known degree-diameter graphs at that size using
the same network equipment and the same number of servers.}}
\label{fig:ev:ddg}
%\vspace{-20pt}
\end{figure}

For our comparisons with the best-known degree-diameter graphs, 
the number of servers we attach to the switches 
was decided such that full-bisection bandwidth was not hit for the
degree-diameter graphs (thus ensuring that we are measuring the
full capacity of degree-diameter graphs.)
Our results, in Fig.~\ref{fig:ev:ddg}, show that the best-known degree-diameter 
graphs do achieve higher throughput than \sys, and thus improve even more over fat-trees. But in the worst of these
comparisons, \sys still achieves $\sim$$91\%$ of the degree-diameter
graph's throughput. While graphs that are optimal for the degree-diameter problem are not
(known to be) provably optimal for our bandwidth optimization problem,
%While optimal degree-diameter graphs are not (known to be) provably optimal for our bandwidth optimization problem, 
these results strongly suggest that \sys's random topology leaves little room for improvement, even with very carefully-optimized topologies.  And what improvement is possible may not be worth the loss of \sys's incremental expandability.
\cut{
%Hex = 0.411; 2DTorus = 0.550; Ring = 0.598; Jellyfish = 0.737
\begin{table}
\begin{center}
\setlength{\tabcolsep}{.7pt}
%\scriptsize
\small
\begin{tabular}{|c|c|c|c|c|}
\hline
\multirow{2}{0.5in}{ } & \multirow{2}{0.6in}{SW $3$D Hex-Torus} & \multirow{2}{0.5in}{SW $2$D Torus} & \multirow{2}{0.5in}{SW Ring} & \multirow{2}{*}{\sys} \\
 & & & & \\ \hline
Throughput & $0.411$ & $0.550$ & $0.598$ & $0.737$ \\ \hline
\end{tabular}
\end{center}
\end{table}
}
\begin{figure}[!t]
%\vspace{-8pt}
\centering
\includegraphics[height=2.6in, angle=270]{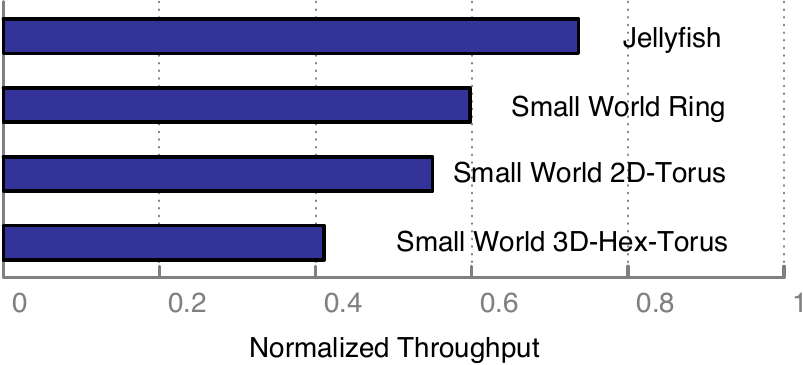}
\caption{\small \em \sys has higher capacity than the (same-equipment) small world data center
topologies~\cite{swdc} built using a ring, a $2$D-Torus, and a $3$D-Hex-Torus as the 
underlying lattice. \newest{Results are averaged over $10$ runs.}}
\label{fig:swdc}
\vspace{-15pt}
\end{figure}

\paragraphb{Throughput vs. small world data centers (SWDC):}
\newx{SWDC~\cite{swdc} proposes a new topology for data centers inspired by a small-world distribution. 
We compare \sys with SWDC using the same degree-$6$ topologies described in the SWDC
paper.} We emulate their $6$-interface server-based design by using switches 
connected with $1$ server and $6$ network ports each. 
We build the three SWDC variants described in~\cite{swdc} at topology
sizes as close to each other as possible (constrained by the lattice
structure underlying these topologies) across sizes we can simulate. 
Thus, we use $484$ switches for \sys, the SWDC-Ring topology, and the 
SWDC-2D-Torus topology; for the SWDC-3D-Hex-Torus, we
use $450$ nodes. (Note that this gives the latter topology an advantage,
because it uses the same degree, but a smaller number of nodes. However,
this is the closest size where that topology is well-formed.) 
\newx{At these sizes, the first three topologies all yielded full throughput, so, to 
distinguish between their capacities, we oversubscribed each topology by 
connecting $2$ servers at each switch instead of just one.}
The results are shown in
Fig.~\ref{fig:swdc}. \sys's throughput is $\sim$$119\%$ of that of the closest
competitor, the ring-based small world topology. \cut{The trend of more
randomness working better is also clearly elucidated through this
experiment.}

\paragraphb{Path Length:} Short path lengths are important to ensure low latency, 
and to minimize network utilization. In this context, we note that the theoretical 
\emph{upper-bound} on the diameter of random regular graphs is fairly small: Bollob\'{a}s and de la Vega~\cite{diabound} 
showed that in almost every $r$-regular graph with $N$ nodes, the diameter
is at most $1 + \lceil \log_{r-1}( (2 + \epsilon)rN \log N ) \rceil$ 
for any $\epsilon > 0$. Thus, the server-to-server diameter is at
most $3 + \lceil \log_{r-1}( (2+\epsilon)rN \log N ) \rceil$. Thus, the path length 
increases logarithmically (base $r$) with the number of nodes in the network.
Given the availability of commodity servers with large port counts, this rate
of increase is very small in practice.

\begin{figure}[!t]
\centering
\includegraphics[width=3.1in]{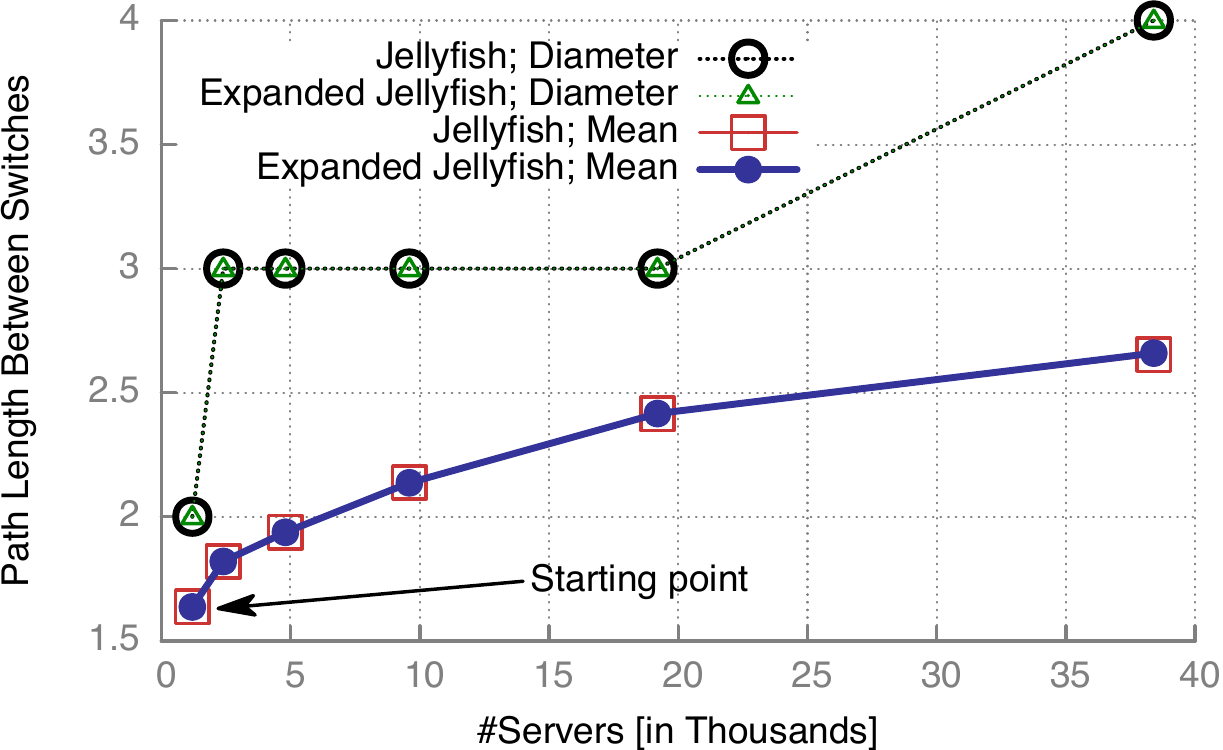}
\caption{\small \em \sys has short paths: Path length versus number of servers, 
with $k=48$ port switches of which $r=36$ connect to other switches and 
$12$ connect to servers. Each data point is derived from $10$ graphs. The diameter
is $\leq$$4$ at these scales. This figure also shows that constructing \sys from
scratch, or using incremental growth yields topologies with very similar path
length characteristics.}
\label{fig:pathlength}
\vspace{-12pt}
\end{figure}

We measured path lengths using an all-pairs shortest-paths algorithm.
The average path length (Fig.~\ref{fig:pathlength}) in \sys is much 
smaller than in the fat-tree\footnote{\newest{Note that the results in Fig.~\ref{fig:pathlength} use
$48$-port switches throughout, meaning that the only point of direct, fair comparison
with a fat-tree is at the largest scale, where \sys still compares favorably against a fat-tree
built using $48$-port switches and $27$,$648$ servers.}}. 
For example, for RRG($3200$, $48$, $36$) 
with $38$,$400$ servers, the average path length between switches is $<$$2.7$ 
(Fig.~\ref{fig:pathlength}), while the fat-tree's average is \newest{$3.71$ at
the smallest size, \cycready{$3.96$ at the size of $27$,$648$ servers.}}
Even though \sys's diameter is $4$ at the largest scale, the $99.99^{\mathrm{th}}$ 
percentile path-length across
$10$ runs did not exceed $3$ for any size in Fig.~\ref{fig:pathlength}.

\vspace{-7pt}
\subsection{Flexibility}
\label{sec:flexibility}
\vspace{-7pt}

\paragraphb{Arbitrary-sized Networks:} Several existing proposals
admit only the construction of interconnects with very coarse
parameters. For instance, a $3$-level fat-tree allows only $k^{3}/4$
servers with $k$ being restricted to the port-count of available 
switches, unless some ports are left unused. 
This is an arbitrary constraint, extraneous to operational 
requirements. In contrast, \sys permits any number of racks 
to be networked efficiently.

\paragraphb{Incremental Expandability:} \sys's construction makes it 
amenable to incremental expansion by adding either
servers and/or network capacity (if not full-bisection
bandwidth already), with increments as small as one rack
or one switch. \sys can be expanded such that rewiring is limited to 
the number of ports being added to the network;
and the desirable properties are maintained: high bandwidth and
short paths at low cost.

As an example, consider an expansion from an RRG($N$, $k$, $r$) topology
to RRG($N+1$, $k$, $r$).  In other words, we are adding one rack of
servers, with its ToR switch $u$, to the existing network. 
We pick a random link $(v,w)$ such that this new ToR switch is not
already connected with either $v$ or $w$, remove it,
and add the two links $(u,v)$ and $(u,w)$, thus using $2$ ports
on $u$. This process is repeated until all ports are filled (or a 
single odd port remains, which could be matched with another free port 
on an existing rack, used for a server, or left free).  This completes 
incorporation of the rack, and can be repeated for as many new racks 
as desired.

\begin{figure}
%\vspace{-8pt}
\centering
\includegraphics[width=3.1in]{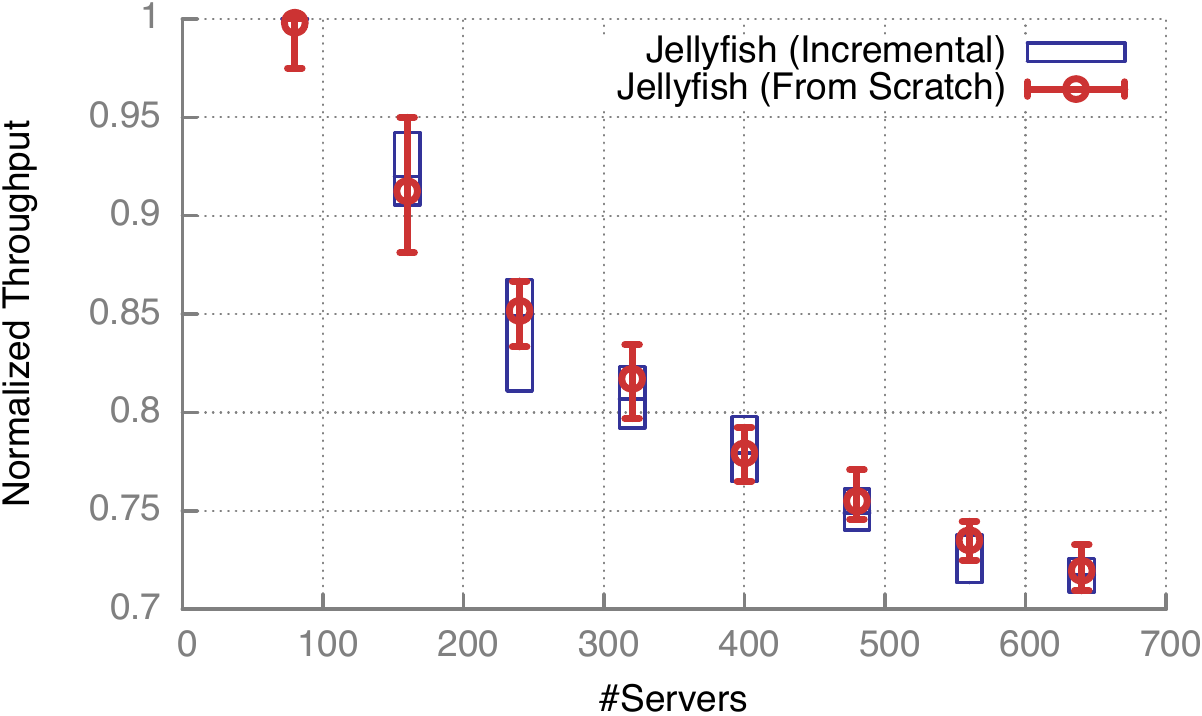}
\caption{\small \em Incrementally constructed \sys has the same
capacity as \sys built from scratch: We built a \sys topology incrementally from $20$
 to $160$ switches in increments of $20$ switches, and compared the                                                  
throughput per server of these incrementally grown topologies to \sys                                                        
topologies built from scratch using our construction routine. The
plot shows the average, minimum and maximum throughput over $20$ runs.
% are shown 
%under optimal routing over random-permutation traffic.
\cut{has $12$ ports, of which $4$ are connected to servers. We compared the 
throughput per server of these incrementally grown topologies to \sys 
topologies built from scratch using our construction routine. Results 
shown are over $20$ runs.}}
\label{fig:expansion}
\vspace{-12pt}
\end{figure}

A similar procedure can be used to expand network capacity for an 
under-provisioned \sys network. In this case, instead of adding a 
rack with servers, we only add the switch, connecting all its ports to 
the network.

\sys also allows for heterogeneous expansion: nothing in the procedure
above requires that the new switches have the same number of ports as the
existing switches. Thus, as new switches with higher port-counts become
available, they can be readily used, either in racks or to augment the
interconnect's bandwidth. There is, of course, the possibility of taking
into account heterogeneity explicitly in the random graph construction
and to improve upon what the vanilla random graph model yields. This
endeavor remains future work for now.

We note that our expansion procedures (like our construction procedure) may not produce uniform-random
RRGs. However, we demonstrate that the path length and capacity 
measurements of topologies we build incrementally match closely with
ones constructed from scratch. Fig.~\ref{fig:pathlength} shows this
comparison for the average path length and diameter where we start with
an RRG with $1$,$200$ servers and expand it incrementally. Fig.~\ref{fig:expansion}
compares the normalized throughput per server under a random permutation traffic 
model for topologies built incrementally against those built from scratch.
The incremental topologies here are built by adding successive increments of $20$ 
switches, and $80$ servers to an initial topology also with $20$ switches and $80$ 
servers. (Throughout this experiment, each switch has $12$ ports, $4$ of which are 
attached to servers.) In each case, the results are close to identical.

\paragraphb{Network capacity under expansion:} \newest{Note that after normalizing by the number of servers $N(k-r)$, the lower bound on \sys's normalized bisection bandwidth (\S\ref{sec:efficiency}) is independent of network size $N$.\cut{, i.e., bound stays constant as the network grows.}} Of course, as $N$ increases with fixed network degree $r$, average path length increases, and therefore, the demand for additional per-server capacity increases\footnote{This discussion also serves as a reminder that bisection-bandwidth, while a good metric of network capacity, is not the same as, say, capacity under worst-case traffic patterns.}. But since path length increases very slowly (as discussed above), bandwidth per server remains high even for relatively large factors of growth. Thus, operators can keep the servers-per-switch ratio constant even under large expansion, with minor bandwidth loss. Adding only switches (without servers) is another avenue for expansion which can preserve or even increase network capacity. Our below comparison with LEGUP uses both forms of expansion.

\begin{figure}[!t]
\centering
\includegraphics[width=3.05in]{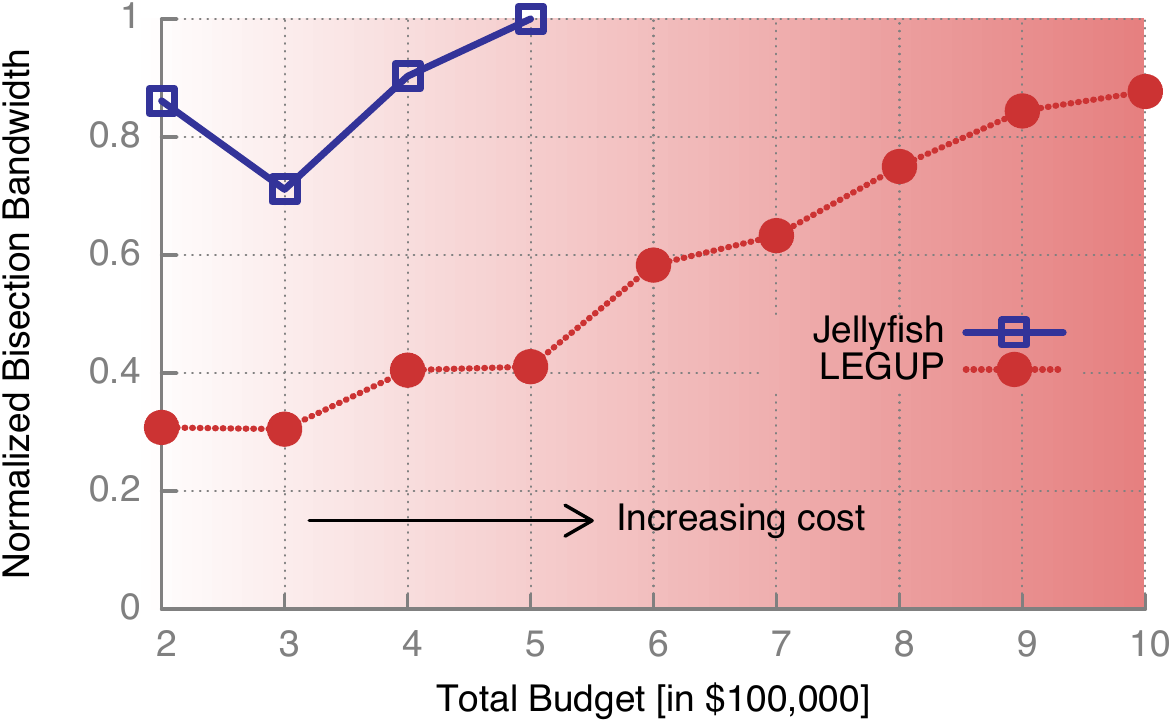}
\caption{\small \em \sys's incremental expansion is substantially more cost-effective than LEGUP's Clos network expansion.  With the same budget for equipment and rewiring at each expansion stage ($x$-axis), \sys obtains significantly higher bisection bandwidth ($y$-axis). Results are averaged over $10$ runs. (The drop in \sys's bisection bandwidth from stage $0$ to $1$ occurs because the number of servers increases in that step.) }
\label{fig:legupcomparison}
\vspace{-12pt}
\end{figure}

\paragraphb{Comparison with LEGUP~\cite{legup}:} While a LEGUP implementation is not publicly available, the authors were kind enough to supply a series of topologies produced by LEGUP. In this expansion arc, there is a budget constraint for the initial network, and for each successive expansion step; within the constraint, LEGUP attempts to maximize network bandwidth, and also may keep some ports free in order to ease expansion in future steps. The initial network is built with $480$ servers and $34$ switches; the first expansion adds $240$ more servers and some switches; and each remaining expansion adds only switches. To build a comparable \sys network, at each expansion step, under the same budget constraints, (using the same cost model for switches, \emph{cabling, and rewiring}) we buy and randomly cable in as many new switches as we can. The number of servers supported is the same as LEGUP at each stage.

LEGUP optimizes for bisection bandwidth, so we compare both LEGUP and \sys on that metric (using code provided by the LEGUP authors~\cite{legup}) rather than on our previous random permutation throughput metric.
The results are shown in Fig.~\ref{fig:legupcomparison}. \sys obtains
substantially higher bisection bandwidth than LEGUP at each stage.  In fact, by stage 2, \sys has achieved higher bisection bandwidth than LEGUP in stage 8, meaning (based on each stage's cost) that \sys builds an equivalent network at cost $60$\% lower than LEGUP.

A minority of these savings is explained by the fact that \sys is more bandwidth-efficient than Clos networks, as exhibited by our earlier comparison with fat-trees. But in addition, LEGUP appears to pay a significant cost to enable it to incrementally-expand a Clos topology; for example, it leaves some ports unused in order to ease expansion in later stages.  We conjecture that to some extent, this greater incremental expansion cost is fundamental to Clos topologies.

\cut{
\begin{figure}[!t]
\centering
\subfigure[]{ \label{fig:inc_fail_rate}\includegraphics[width=2.8in]{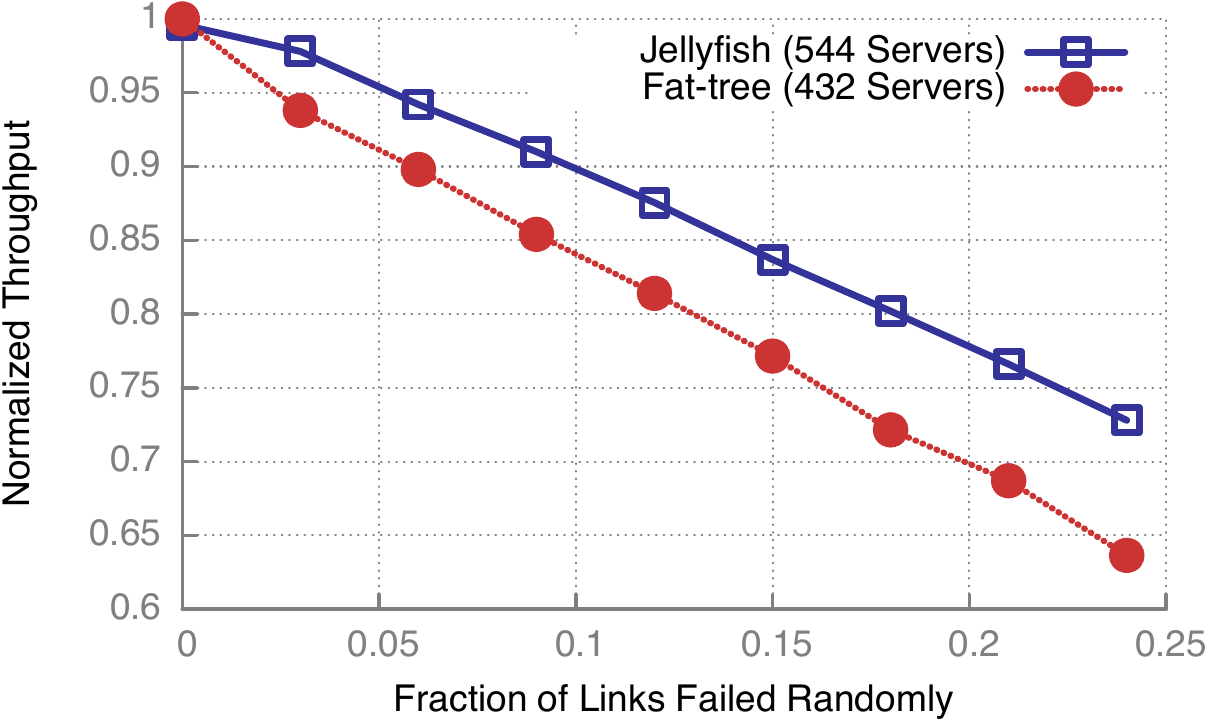}}
\vspace{-12pt}
\subfigure[]{ \label{fig:fail_inc_size}\includegraphics[width=2.8in]{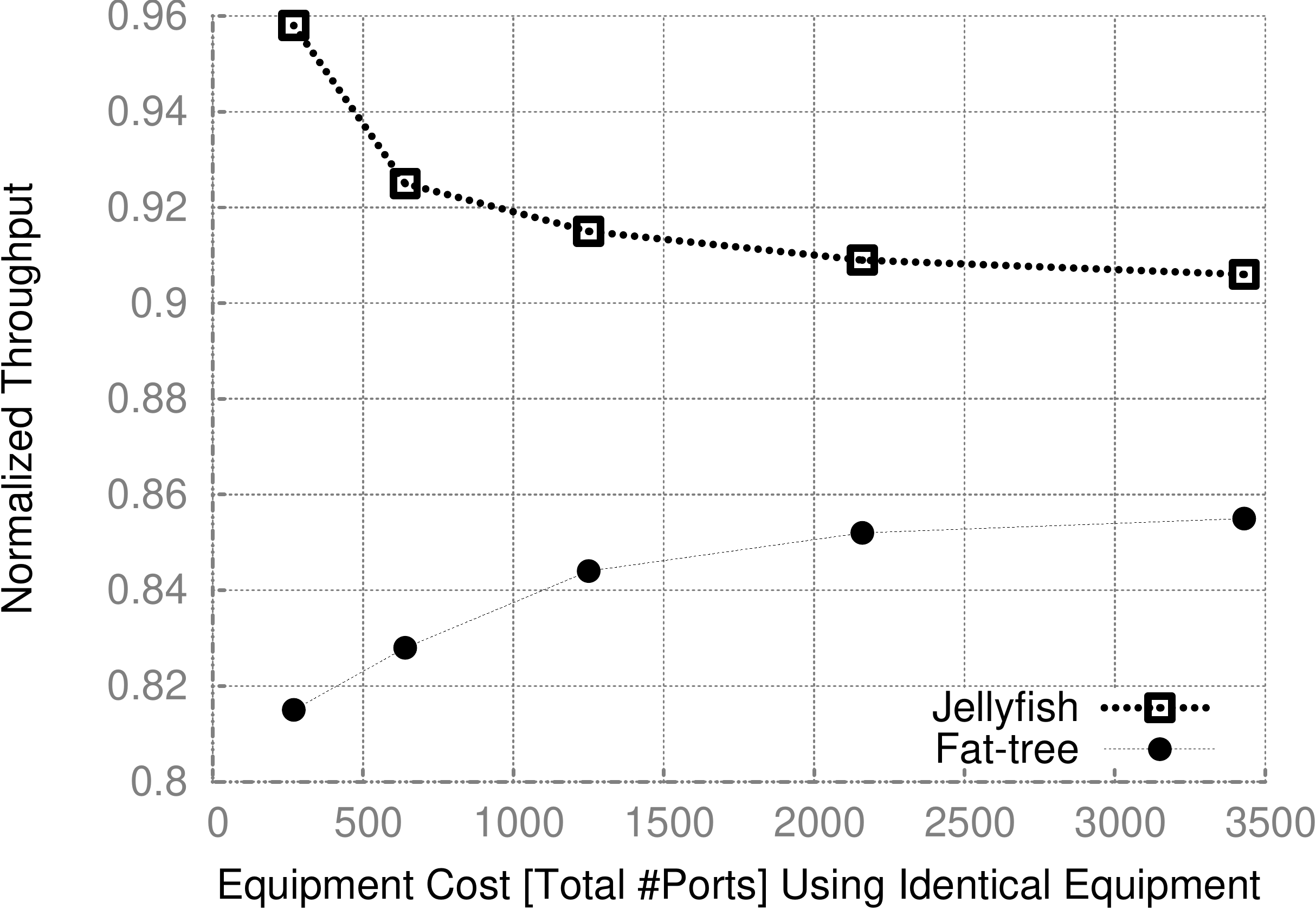}}
%\subfigure[]{ \label{fig:plall}\includegraphics[width=2.2in]{figures/plall}}
\caption{\small \em \sys is highly resilient to failures: a) Normalized
throughput per server decreases more gracefully for \sys than for a same-equipment
fat-tree as the percentage of failed links increases. Note that the y-axis 
starts at $60\%$ throughput; both topologies are highly resilient to failures.
b) With a fixed link failure rate ($9\%$), across increasing topology size, \sys 
maintains the resilience advantage over the fat-tree. We note that the particular
topology used for the experiment in a) is the one with equipment cost $2$,$160$,
i.e. the last-to-second point in the plot in b). Results are averaged over $10$
runs.}
\label{fig:resilience}
\vspace{-12pt}
\end{figure}
}

\begin{figure}[!t]
\centering
\includegraphics[width=3.1in]{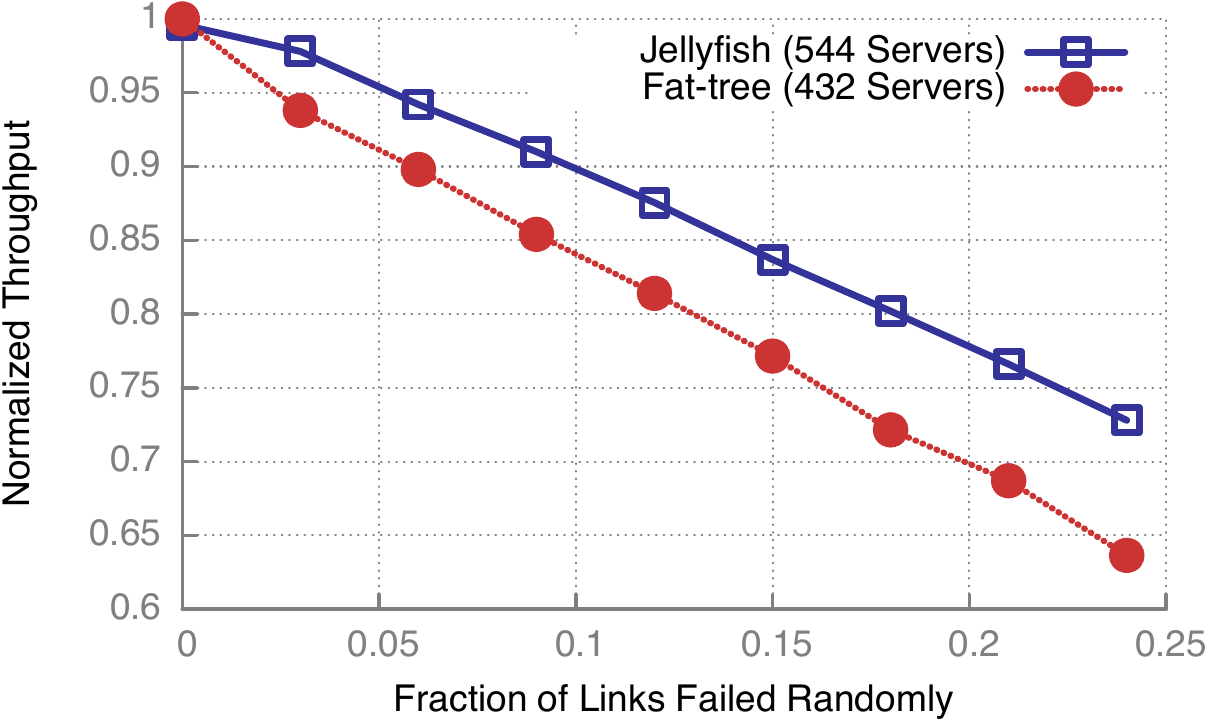}
\caption{\small \em \sys is highly resilient to failures: Normalized
throughput per server decreases more gracefully for \sys than for a same-equipment
fat-tree as the percentage of failed links increases. Note that the y-axis 
starts at $60\%$ throughput; both topologies are highly resilient to failures.
\cut{b) With a fixed link failure rate ($9\%$), across increasing topology size, \sys 
maintains the resilience advantage over the fat-tree. We note that the particular
topology used for the experiment in a) is the one with equipment cost $2$,$160$,
i.e. the last-to-second point in the plot in b). Results are averaged over $10$
runs.}}
\label{fig:resilience}
\vspace{-12pt}
\end{figure}

\subsection{Failure Resilience} 
\label{sec:failure}
\ankitb{\sys provides good
path redundancy; in particular, an $r$-regular random graph 
is almost surely $r$-connected~\cite{kconnected}.} Also, the random topology maintains 
its (lack of) structure in the face of link or node 
failures -- a random graph topology with a few failures is just another 
random graph topology of slightly smaller size, with a few unmatched ports 
on some switches.

\newest{Fig.~\ref{fig:resilience} shows that the \sys topology is even more
resilient than the same-equipment fat-tree (which itself is no weakling). \cut{Fig.~\ref{fig:inc_fail_rate}
compares a fat-tree and a same-equipment \sys topology as the fraction 
of links failed (uniformly at random) increases. Fig.~\ref{fig:fail_inc_size}
compares the failure resilience of the fat-tree and \sys as the size of
the topology increases \newx{while keeping a fixed failure rate of $9\%$}.} Note that
the comparison features a fat-tree with fewer servers, but the same
cost. This is to justify \sys's claim of supporting a larger number
of servers using the same equipment as the fat-tree, in terms of 
capacity, path length, and \emph{resilience} simultaneously.
}
%\begin{figure}
%\vspace{-12pt}
%\centering
%\includegraphics[width=3in]{figures/increasing_fail}
%\caption{\small \em Flow-level simulation. \sys versus fat-tree.
%Per-server throughput decreases more gracefully for a (same switching equipment) \sys topology 
%than the fat-tree (which itself is very resilient). Not only does the
%same-equipment \sys topology support more servers, it is also more failure
%resilient.}
%\label{fig:inc_fail_rate}
%\end{figure}
%
%
%\begin{figure}[!t]
%\vspace{-12pt}
%\centering
%\includegraphics[width=3in]{figures/fail_increasing_size}
%\caption{\small \em Flow-level simulation. Failure resilience of \sys versus fat-tree
%as the topology-size increases (with a fixed failure rate of 9\% of links). }
%\label{fig:ev:fat}
%\end{figure}
%\paragraphb{Resilience to miswirings:} Intuitively, a few miswirings
%just make it a different random graph than the one `planned'. Unless
%these are very numerous and biased,
%\sys will continue to preserve its properties. 

%\ankitb{Miswirings in other
%topologies can downgrade performance~\cite{miswirings}.} 

%\vspace{-7pt}

\section{Routing \& Congestion Control}
\label{sec:routing}

While the above experiments establish that \sys topologies have high capacity, it remains unclear 
whether this potential can be realized in real networks. There are two layers which can 
affect performance in real deployments: routing and congestion control. \newest{In our experiments with various 
combinations of routing and congestion control for \sys (\S\ref{sec:routing:ecmpnotenough}), we find 
that standard ECMP does not provide enough path diversity for \newestx{\sys, and to utilize the entire capacity we need to also use longer paths.
We then provide in-depth results for \sys's throughput and fairness using the best setting found earlier---$k$-shortest paths and multipath TCP (\S\ref{sec:routing:simulation}).}
%Our later results on \sys's throughput and fairness (\S\ref{sec:routing:simulation}) use the best setting we 
%found -- multipath TCP over $k$-shortest paths.
Finally, we discuss practical strategies for deploying $k$-shortest-path routing (\S\ref{sec:routing:deployment}).}

\subsection{ECMP is not enough}
\label{sec:routing:ecmpnotenough}

\begin{figure}[!t]
\centering
\includegraphics[width=3.1in]{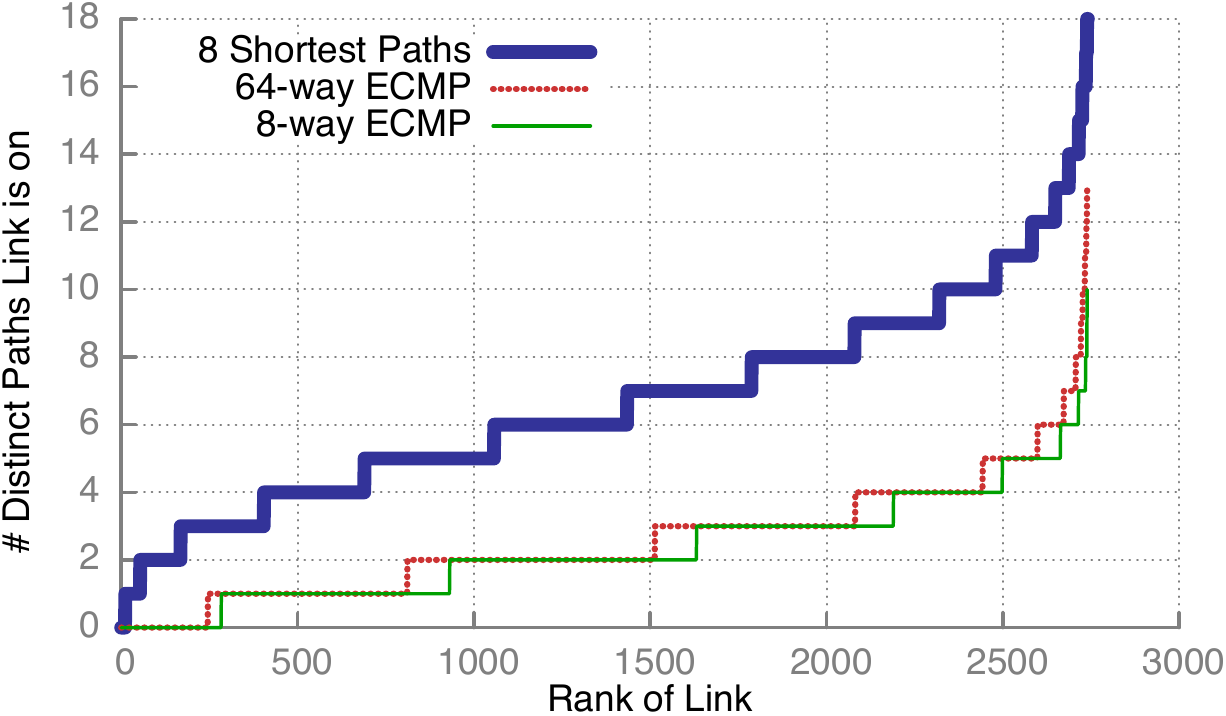}
\caption{\small \em ECMP does not provide path diversity for \sys: Inter-switch link's path 
count in ECMP and $k$-shortest-path routing for random permutation traffic at the server-level on a typical 
Jellyfish of $686$ servers (built using the same equipment as a fat-tree supports $686$ servers). 
For each link, we count the number of distinct paths it is on. Each network cable is considered 
as two links, one for each direction.}
\label{fig:pathdiversity}
%\vspace{-12pt}                                                                                                               
\end{figure}

\paragraphb{Evaluation methodology:} We use the simulator 
developed by the MPTCP authors for both \sys and fat-tree. 
For routing, we \newestx{test: (a)} ECMP (equal cost multipath routing; \newest{We used $8$-way ECMP,
but $64$-way ECMP does not perform much better, see Fig.~\ref{fig:pathdiversity}}), a standard strategy 
to distribute flows over shortest paths; and \newestx{(b)} $k$-shortest paths routing, which could be useful for \sys because it can utilize longer-than-shortest paths. For $k$-shortest paths, we use Yen's Loopless-Path Ranking algorithm~\cite{yen, yenimpl} with $k=8$ paths. For congestion control, we test standard TCP ($1$ or $8$ flows per server pair) and  
the recently proposed multipath TCP (MPTCP)~\cite{mptcp}, using the recommended value of $8$ MPTCP subflows.
The traffic model continues to be a random permutation at the server-level, and as before, 
for the fat-tree comparisons, we build \sys using the same switching equipment as the fat-tree.

\paragraphb{Summary of results:} 
%\newest{In Table~\ref{tbl:routing}, we show 
%average per server throughput (as a percentage of server's NIC rate)
%for both topologies under different routing and congestion control schemes.} 
\newestx{Table~\ref{tbl:routing} shows the average per server throughput 
as a percentage of the servers' NIC rate 
for two sample \sys and fat-tree topologies under different routing and load balancing schemes.} 
We make two observations: (1) ECMP performs poorly for \sys, not providing 
enough path diversity. \cycready{For random permutation traffic, Fig.~\ref{fig:pathdiversity} shows that 
about $55\%$ of links are used by no more than $2$ paths under ECMP; while for $8$-shortest path routing, the number is $6\%$.} 
Thus we need to make use of $k$-shortest paths. (2) Once we use $k$-shortest paths, each congestion control protocol works as least as well for \sys as for the fat-tree. 

\newestx{The results of Table~\ref{tbl:routing} depend on the oversubscription level of the network. In this context, we attempt to match fat-tree's performance given the routing and congestion control inefficiencies. We found that \sys's advantage slightly reduces in this context compared to using idealized routing as before:} In comparison to the same-equipment fat-tree ($686$ servers),
now we can support, at same or higher performance, $780$ servers (\textit{i.e.}, $13.7\%$ more
that the fat-tree) with TCP, and $805$ servers ($17.3\%$ more) with MPTCP. With ideal routing and congestion control,
\sys could support $874$ servers ($27.4\%$ more). 
%While the gains from \sys are reduced, they are still sizable. 
\fixme{removed phrase}
\newestx{However}, as we show quantitatively in \S\ref{sec:routing:simulation},
\sys's advantage improves significantly with scale. At the largest scale we could simulate, 
\sys supports $3$,$330$ servers to the fat-tree's $2$,$662$ --- a $>25\%$ improvement 
(after accounting for routing and congestion control inefficiencies).

%Note that routing and congestion control inefficiencies have reduced
%\sys's advantage a little: In comparison to the same-equipment fat-tree ($686$ servers),
%now we can support, at same or higher performance, $780$ servers (\textit{i.e.}, $13.7\%$ more
%that the fat-tree) with TCP, and $805$ servers ($17.3\%$ more) with MPTCP. With ideal routing and congestion control,
%\sys could support $874$ servers ($27.4\%$ more). While the gains from \sys are reduced,
%they are still sizable. Further, as we show quantitatively in \S\ref{sec:routing:simulation},
%\sys's advantage improves significantly with scale. At the largest scale we could simulate, 
%\sys supports $3$,$330$ servers to the fat-tree's $2$,$662$ -- a $>25\%$ improvement 
%(after accounting for routing and congestion control inefficiencies).
%%At the largest scale we could simulate (using MPTCP with $8$-shortest paths), 
%%\sys supports $3$,$330$ servers to the fat-tree's $2$,$662$ -- a $>$$25\%$ improvement.
%%There is also more opportunity for work on better routing and congestion control over \sys.

\begin{table}[t]
\begin{center}
\setlength{\tabcolsep}{.7pt}
%\scriptsize
\small
%%%%%%%%%  OLDEST VERSION %%%%%%%%%%%
%\begin{tabular}{c|c|c|c|}
%\cline{2-4}
%          & \multicolumn{2}{|c|}{\sys ($780$ svrs)}   & Fat-tree ($686$ svrs)\\ \cline{2-4}
%           & $8$-shortest paths         & ECMP  & ECMP \\ \hline
%\multicolumn{1}{|c|} {TCP $1$ flow}      & $48.3\%$      & $57.9\%$   & $48.0\%$  \\ \hline
%\multicolumn{1}{|c|} {TCP $8$ flows}     & $92.3\%$      & $73.9\%$   & $92.2\%$  \\ \hline
%\multicolumn{1}{|c|} {MPTCP $8$ subflows}& $95.1\%$      & $76.4\%$   & $93.6\%$  \\ \hline
%\end{tabular}
%%%%%%%%%  OLD VERSION %%%%%%%%%%%
%\begin{tabular}{c|c|c|c|}
%\cline{2-4}
%          & 	Fat-tree ($686$ svrs) &  \multicolumn{2}{|c|}{\sys ($780$ svrs)}\\ \cline{2-4}
%          			      &    ECMP 	&  ECMP      & $8$-shortest paths      \\ \hline
%\multicolumn{1}{|c|} {TCP $1$ flow}   &    $48.0\%$     & $57.9\%$   & $48.3\%$ \\ \hline
%\multicolumn{1}{|c|} {TCP $8$ flows}  &    $92.2\%$     & $73.9\%$   & $92.3\%$   \\ \hline
%\multicolumn{1}{|c|} {MPTCP $8$ subflows}& $93.6\%$  	& $76.4\%$   & $95.1\%$     \\ \hline
%\end{tabular}
%%%%%%%%%  NEW VERSION %%%%%%%%%%%
\begin{tabular}{c|c|c|c}
\hline 
\multicolumn{1}{|c|}{Congestion} & Fat-tree ($686$ svrs) & \multicolumn{2}{c|}{\sys ($780$ svrs)}\tabularnewline
\multicolumn{1}{|c|}{control} & ECMP & ECMP & \multicolumn{1}{c|}{$8$-shortest paths}\tabularnewline
\hline 
\hline 
TCP 1 flow & $48.0\%$ & $57.9\%$ & $48.3\%$\tabularnewline
TCP 8 flows & $92.2\%$ & $73.9\%$ & $92.3\%$\tabularnewline
MPTCP 8 subflows & $93.6\%$ & $76.4\%$ & $95.1\%$\tabularnewline
\hline 
\end{tabular}
\caption{\small \em Packet simulation results for different routing and congestion control protocols
for \sys ($780$ servers) and a same-equipment fat-tree ($686$ servers). Results show the normalized per server
average throughput \newestx{as a percentage of servers' NIC rate}
%(note that \sys has $\mathbf{13.7}\%$ more servers than fat-tree) 
%\fixme{removed for space}
over $5$ runs.
We did not simulate the fat-tree with $8$-shortest paths because ECMP is strictly better, and easier
to implement in practice for the fat-tree.}
\label{tbl:routing}
\end{center}
\end{table}

\subsection{$k$-Shortest-Paths With MPTCP}
\label{sec:routing:simulation}

\begin{figure}[!t]
\centering
\includegraphics[width=2.8in]{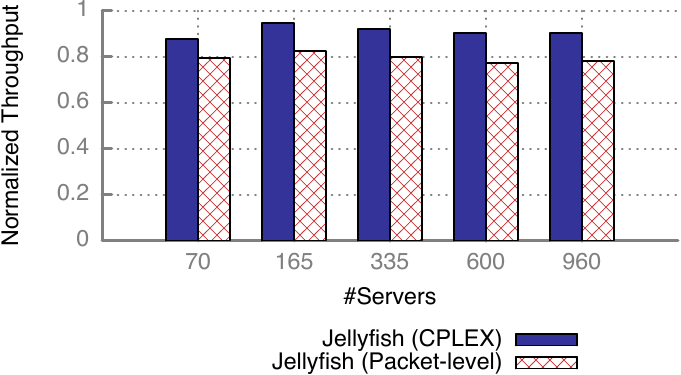}
\caption{\small \em Simple $k$-shortest path forwarding with MPTCP exploits \sys's high
capacity well: We compare the throughput using the same \sys topology
with both optimal routing, and our simple routing mechanism using MPTCP,
which results in throughput between $86\% - 90\%$ of the optimal routing
in each case. Results are averaged over $10$ runs.}
\label{fig:flow_vs_packet}
\vspace{-8pt}
\end{figure}

\newest{The above results demonstrate using one representative set of topologies
that using $k$-shortest paths with MPTCP yields higher performance than ECMP/TCP. 
In this section we measure the efficiency of $k$-shortest path routing 
with MPTCP congestion control against the optimal performance (presented in \S\ref{sec:researchopps}),
and later make comparisons against fat-trees at various sizes.}

\paragraphb{Routing and Congestion Control Efficiency:} 
The result in Fig.~\ref{fig:flow_vs_packet} shows the gap between
the optimum performance, and the performance realized with routing
and congestion control inefficiencies. At each size, we use the 
same slightly oversubscribed\footnote{\newx{An undersubscribed network may simply show $100$\% throughput, masking some of 
the routing and transport inefficiency.}} \sys topology for both 
setups. In the worst of these
comparisons, \sys's packet-level throughput is at $\sim$$86\%$ of the CPLEX
optimal throughput. 
(In comparison, the fat-tree's throughput under MPTCP/ECMP is $93$-$95\%$ 
of its optimum.) 
%\fixme{removed parenthetical}
There is a possibility that this gap can be 
closed using smarter routing schemes, but nevertheless, as we discuss below, 
\sys maintains most of its advantage over the fat-tree in
terms of the number of servers supported at the the same throughput.

\begin{figure}[!t]
\centering
\includegraphics[width=3.1in]{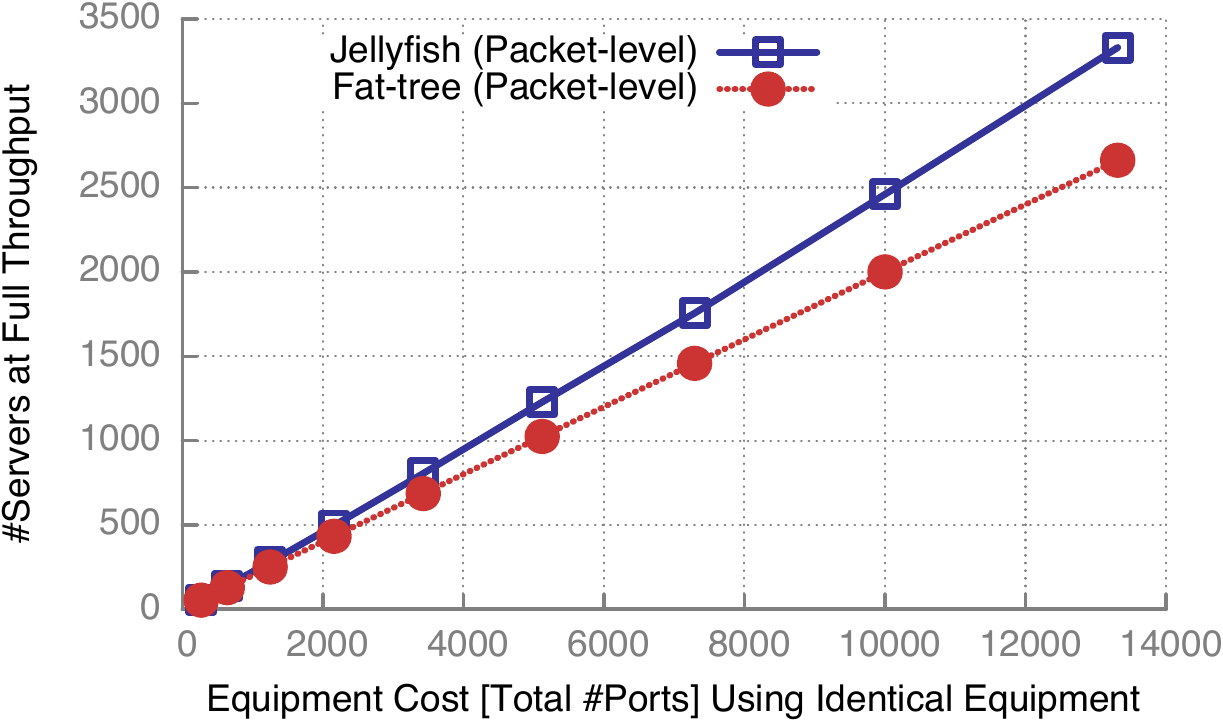}
\caption{\small \em \sys supports a larger number of servers ($>$$25\%$ at the largest
scale shown, with an increasing trend) than the
same-equipment fat-tree at the same (or higher) throughput, even with 
inefficiencies of routing and congestion control accounted for.
\cut{We measure the number of servers the fat-tree and \sys can support using the same
switching equipment for the same per-server throughput. 
Using packet-level simulation, we show that \sys still supports a significantly larger number of 
servers than the fat-tree, for the same per-server throughput as the fat-tree.}
Results are averages over $20$ runs for topologies smaller than $1$,$400$ servers,
and averages over $10$ runs for larger topologies.}
\label{fig:fat_thput_htsim}
%\vspace{-12pt}                                                                                                               
\end{figure}

\paragraphb{Fat-tree Throughput Comparison:} To compare \sys's 
performance against the fat-tree, we first find the average per-server 
throughput a fat-tree yields in the packet simulation.
We then find \newestx{(using binary search)} the number of servers
for which the average per-server throughput for the comparable \sys 
topology is either the same, or higher than the fat-tree; \newestx{this is the same methodology applied for Table~\ref{tbl:routing}}. We repeat 
this exercise for several fat-tree sizes. The results (Fig.~\ref{fig:fat_thput_htsim}) are similar
to those in Fig.~\ref{fig:fat_cplex}, although the gains of \sys
are reduced marginally due to routing and congestion control inefficiencies. 
Even so, at the maximum scale of our experiment, \sys supports $25\%$ more servers than the fat-tree
($3$,$330$ in \sys, versus $2$,$662$ for the fat-tree). We note
however, that even at smaller scale (for instance, $496$ servers in \sys, 
to $432$ servers in the fat-tree) the improvement can be as large
as $\sim$$15\%$.

\begin{figure}[!t]                                                                                                           
\centering                                                                                                                   
\includegraphics[width=3.1in]{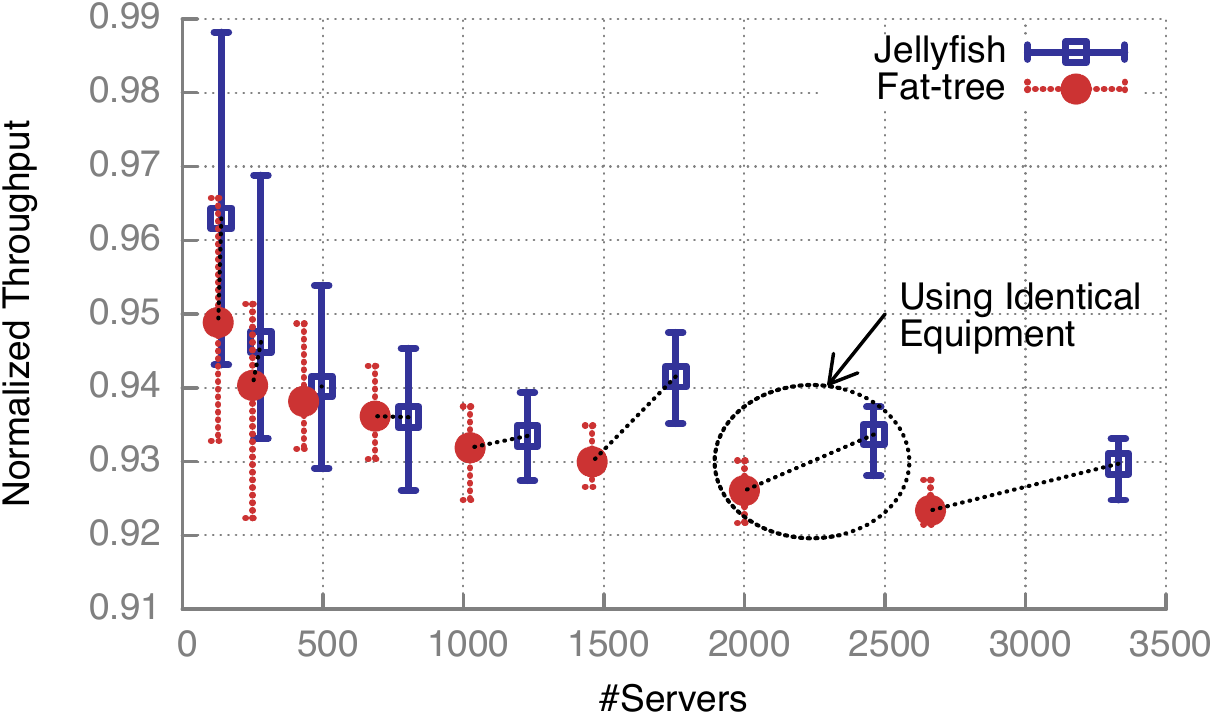}                                                                      
\caption{\small \em The packet simulation's throughput results 
for \sys show similar stability as the fat-tree. (Note that the y-axis starts at $91\%$
throughput.) Average, minimum and maximum throughput-per-server values
are shown. The data plotted is from the same experiment as 
Fig.~\ref{fig:fat_thput_htsim}.
\sys has the same or higher average throughput as the fat-tree while 
supporting a larger number of servers. Each \sys data-point uses equipment identical to the closest fat-tree
data-point to its left (as highlighted in one example).}
\label{fig:stability}                                                                                                        
\end{figure}

\begin{figure}[t]
%\vspace{-12pt}
\centering
\includegraphics[width=3.1in]{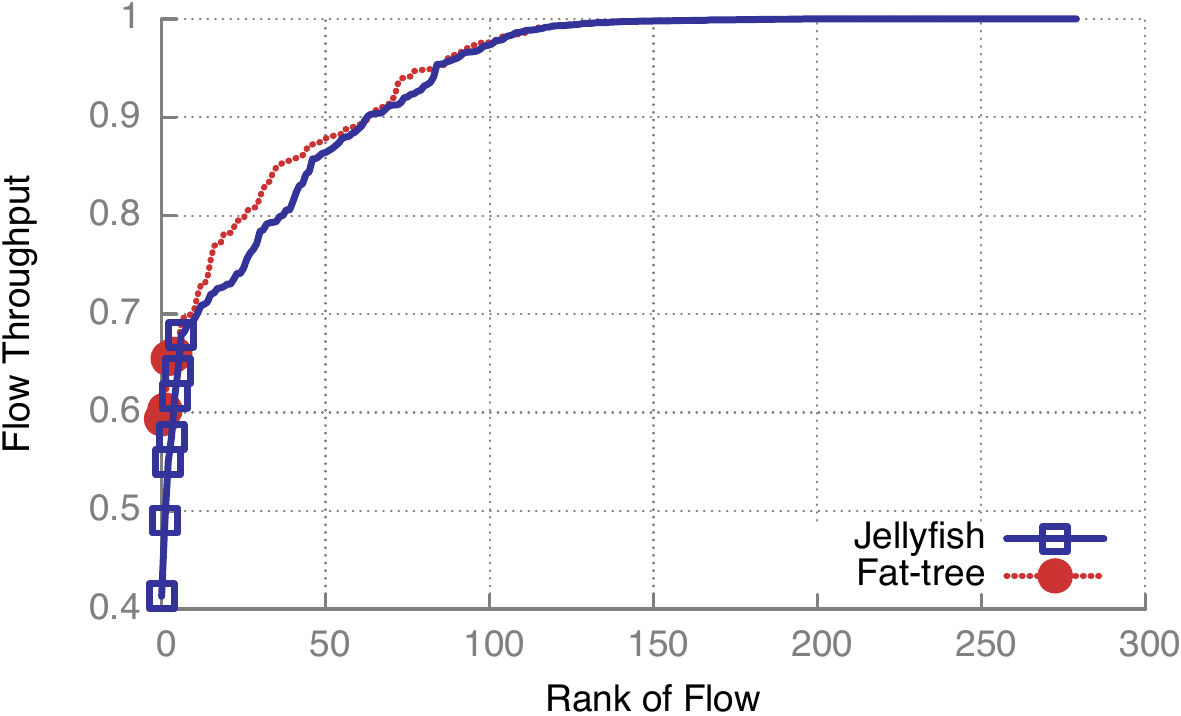}
\caption{\small \em Both \sys and the fat-tree show good flow-fairness:
The distribution of normalized flow throughputs in \sys and fat-tree is
shown for one typical run. After the few outliers (shown with points), the plot is virtually 
continuous (the line). Note that \sys has more flows because it supports a
higher number of servers (at same or higher \emph{per-server} throughput).
Jain's fairness index for both topologies is $\sim$$99\%$.}
\label{fig:fat_fairness}
\vspace{-12pt}                                                                                                               
\end{figure}

We show in Fig.~\ref{fig:stability}, the stability of our
experiments by plotting the average, minimum and maximum throughput for
both \sys and the fat-tree at each size, over $20$ runs \newest{(varying both 
topologies and traffic)} for small sizes
and $10$ runs for sizes $>$$1$,$400$ servers.
%We use a smaller  number of runs for the larger topologies in the interest of running time. 
%We note that results were more 
%stable for large topologies, with a standard deviation of $<$$0.5\%$ of the mean 
%versus $<$$2\%$ of the mean in each case for smaller topologies.
%\fixme{removed text for space}

\paragraphb{Fairness:} We evaluate how flow-fair the routing and congestion control
is in \sys. We use the packet simulator to measure each flow's
throughput in both topologies and show in Fig.~\ref{fig:fat_fairness}, the normalized
throughput per flow in increasing order. Note that \sys has a larger
number of flows because we make all comparisons using the same network
equipment and the larger number of servers supported by \sys. Both the
topologies have similarly good fairness; Jain's fairness
index~\cite{jainfairness} over these flow throughput values 
for both topologies: $0.991$ for the fat-tree and $0.988$ for \sys.

\subsection{Implementing $k$-Shortest-Path Routing}
\label{sec:routing:deployment}
\cut{Since \sys requires more diverse forwarding paths than ECMP provides, }
In this section, we discuss practical possibilities for implementing $k$-shortest-paths routing. 
For this, each switch needs to maintain a routing table containing for each
other switch, $k$ shortest paths. 

%Note that a few thousand switches
%can support several tens of thousands of servers, so routing table
%sizes are unlikely to be a problem. 
%\fixme{I don't buy this, need reference/clarification or remove}
%\cut{We propose the following plausible 
%approaches for enabling $k$-shortest path forwarding.}

\paragraphb{OpenFlow~\cite{openflow}:} OpenFlow switches can match end-to-end connections to routing rules, 
and can be used for routing flows along pre-computed $k$-shortest paths. Recently, Devoflow~\cite{devoflow} 
showed that OpenFlow rules can be augmented with a small set of local routing actions for
randomly distributing load over allowed paths, without invoking the OpenFlow controller.

\paragraphb{SPAIN~\cite{spain}:} SPAIN allows multipath routing
by using VLAN support in commodity off-the-shelf switches. Given a set of pre-computed paths, SPAIN merges these 
paths into multiple trees, each of which is mapped to a separate VLAN. SPAIN supports 
arbitrary topologies, and can enable use of $k$-shortest path routing in \sys.

\paragraphb{MPLS~\cite{mpls}:} One could set up MPLS tunnels between 
switches such that all the pre-computed $k$-shortest paths between a switch pair are
configured to have the same cost. This would allow switches to perform standard
equal-cost load balancing across paths.

\section{Physical Construction and Cabling}
\label{sec:cabling}

%Chi-Yao: I moved these to section 6.1
%We envision \sys cabling to be performed by using a blueprint 
%automatically generated based on the topology and the physical layout of the data center. 
%This blueprint is handed to workers to connect cables mechanically. 
%With this in mind, in this section, we discuss the following key considerations for wiring data centers with \sys:

Key considerations in data center wiring include:
\begin{itemize}
\item {\bf Number of cables:} Each cable represents both a material and a labor
cost.
\item {\bf Length of cables:} The cable price/meter is $\$5$-$6$ for 
both electrical and optical cables, but the cost of an optical 
transceiver can be close to $\$200$~\cite{tamingmonster}.
We limit our interest in cable length to whether or not a cable is
short enough, i.e., $<$$10$ meters in length~\cite{dragonfly, helios}, 
for use of an electrical cable.
%\luc{check above, did not understand previous paragraph}
\item {\bf Cabling complexity:} Will \sys awaken the dread spaghetti monster?
Complex and irregular cabling layouts may be hard to wire and thus susceptible to
more wiring errors. We will consider whether this is a significant
factor. In addition, we attempt to design layouts that result in 
aggregation of cables in bundles, in order to reduce manual effort 
(and hence, expense) for wiring.
\end{itemize}

In the rest of this section, we first address a common concern across data center deployments: handling wiring errors (\S\ref{subsec:miswiring}). 
We then investigate cabling \sys in two deployment scenarios, using the above metrics (number, length and complexity of cabling) to compare against cabling a fat-tree network. The first deployment scenario is represented by small clusters ($\sim$$1$,$000$ servers); in this category we also include the intra-container clusters for `Container Data Centers' (CDC)\footnote{As early as $2006$, The Sun Blackbox~\cite{sunblackbox}
promoted the idea of using shipping containers for data centers. 
There are also new products in the market exploiting 
similar physical design ideas~\cite{rackable, icecubeair, hp-ecopod}.} (\S\ref{sec:smalldcn}).
The second deployment scenario is represented by massive-scale data centers (\S\ref{sec:massive}). In this paper we only analyze massive data centers built using containers, leaving more traditional data center layouts to future work.\footnote{The use of container-based data centers seems to be an industry trend, with several players, Google and Microsoft included, already having container-based deployments~\cite{helios}.}

\subsection{Handling Wiring Errors} 
\label{subsec:miswiring}

We envision \sys cabling being performed using a blueprint automatically generated based on the 
topology and the physical data center layout. 
The blueprint is then handed to workers to connect cables \newest{manually}.

While some human errors are inevitable in cabling, these are easy to detect and fix. 
Given \sys's sloppy topology, 
a small number of miswirings may not even require fixing in many cases. 
Nevertheless, we argue that fixing miswirings is relatively inexpensive.
For example, the labor cost of 
cabling is estimated at $\sim$$10\%$ of total cabling cost~\cite{tamingmonster}. 
With a pessimistic estimate where the total cabling cost is $50$\% of the network cost,
the cost of fixing (for example) $10\%$ miswirings would thus be just $0.5\%$ of the network cost. 
We note that wiring errors can be detected using a link-layer discovery protocol~\cite{lldp}. 
%and checking the resulting cabling against a computer-generated connection plan.

%We note that running a link-layer topology discovery protocol~\cite{lldp} 
%yields enough information to check the resulting cabling against a
%computer-generated connection plan and detect errors.

\subsection{Small Clusters and CDCs}
\label{sec:smalldcn}

Small clusters and CDCs form a significant
section of the market for data centers, and thus merit separate
consideration. In a $2011$ survey~\cite{dcnsurvey} of $300$ US 
enterprises (with revenues ranging from $\$1$B-$\$40$B)
which operate data centers, $57\%$ of data centers occupy
between $5$,$000$ and $15$,$000$ square feet; and $75\%$
have a power load $<$$2$MW, implying that these data centers
house a few thousand servers~\cite{rewire}. As our 
results in \S\ref{sec:efficiency} show, even at a few hundred
servers, cost-efficiency gains from \sys can be significant ($\sim$$20\%$
at $1$,$000$ servers). Thus, it is useful to deploy {\sys} in these scenarios.

We propose a cabling optimization (along similar lines as the one 
proposed in~\cite{fattree}). The key observation is that in a 
high-capacity \sys topology, there are more than 
twice as many cables running between switches than from servers to switches.
Thus, placing all the switches in close proximity to each other reduces cable 
length, as well as manual labor. This also simplifies the small amounts of 
rewiring necessary for incremental expansion, or for fixing wiring errors.

\paragraphb{Number of cables:} Requiring fewer network switches 
for the same server pool also implies requiring fewer
cables ($15-20\%$ depending on scale) than a fat-tree. 
This also implies that there is more room (and budget) for 
packing more servers in the same floor space.

\paragraphb{Length of cables:} For small clusters, and inside
CDC containers using the above optimization, cable 
lengths are short enough for electrical cables without repeaters. 

\paragraphb{Complexity:} For a few thousand servers, space equivalent
to $3$-$5$ standard racks can accommodate the switches needed
for a full bisection bandwidth network (using available $64$-port switches). 
These racks can be placed at the physical center 
of the data center, with aggregate cable bundles running between 
them. From this `switch-cluster', aggregate cables can be run to each 
server-rack. With this plan, manual cabling is fairly simple.
Thus, the nightmare cable-mess image a random graph network may bring 
to mind is, at best, alarmist.

A unique possibility allowed by the assembly-line
nature of CDCs, is that of fabricating a random-connect \emph{patch panel} 
such that workers only plug cables from the switches into the panel in a regular 
easy-to-wire pattern, and the panel's internal design encodes the random 
interconnect. This could greatly accelerate manual cabling.

Whether or not a patch panel is used, the problems of layout and wiring need 
to be solved only \emph{once} at design time for CDCs. With a standard layout 
and construction, building automated tools for verifying and detecting 
miswirings is also a one-time exercise. Thus, the cost of any additional
complexity introduced by \sys would be amortized over the production of many
containers.

%\cut{
%\begin{figure}
%\centering
%\includegraphics[width=2.5in]{figures/wiring}
%\vspace{-8pt}
%\caption{\small \em Physical layout and incremental expansion. \ankitx{Modify 
%figure!!}}
%\label{fig:wiring}
%\vspace{-12pt}
%\end{figure}
%}

\paragraphb{Cabling under expansion:} Small \sys clusters can be expanded 
by leaving enough space near the `switch-cluster' for adding switches
 as servers are added at the periphery of the network. In case
no existing switch-cluster has room for additional switches, a new
cluster can be started. Cable aggregates run from this new switch-cluster 
to all new server-racks and to all other switch-clusters.
We note that for this to work with only electrical cabling, the switch-clusters
need to be placed within $10$ meters of each other as well as the servers.
Given the constraints the support infrastructure already places on such 
facilities, we do not expect this to be a significant issue. 

\cut{An example of such an expansion is shown in Fig.~\ref{fig:wiring}.}

As discussed before, the \sys expansion procedure requires 
a small amount of rewiring.
%a sequence of edge swaps. 
%After an automated computation of the network cables that
%need to be moved and new ones that need attachment, these can be run
%parallel to existing cable bundles, or in the case of new switch-clusters,
%new cable aggregates can be started. 
The addition of every two network ports requires two cables to be moved
(or equivalently, one old cable to be disconnected and two new cables to be added), 
since each new port will be connected to an existing port.
%(one end of an existing cable is connected to one of
%the two new ports, and a new cable is connects the orphaned attachment 
%point of the old cable to the second new port). 
The cables that need to be disconnected and the new cables that need to be attached 
can be automatically identified.
Note that in the
`switch-cluster' configuration, all this activity happens at one 
location (or with multiple clusters, only between these clusters).
The only cables not at the switch-cluster are the ones between the
new switch and the servers attached to it (if any). This is just 
\emph{one} cable aggregate.

We note that the CDC usage 
may or may not be geared towards incremental expansion.
\cut{There are certainly products in the market which bring the modularity
idea containers brought to mega-data centers, to containers themselves
-- allowing gradual build-up of a container with smaller 
container-modules~\cite{icecubeair}. The more common scenario, however, appears to be standard, 
packed containers,}
Here the chief utility of \sys is its efficiency and
reliability. \cut{Nevertheless, \sys expansion ideas apply to the modular 
containers in similar fashion to their application to
small data centers. If patch panels are used in CDCs as suggested, then
they play a role similar to the switch-cluster in the small data center:
most rewiring can be completed at the patch panel.}

\subsection{\sys in Massive-Scale Data Centers}
\label{sec:massive}
We now consider massive scale data centers built by connecting
together multiple containers of the type described above.
%Inside the containers, the same arguments for number, length, and 
%complexity of cables apply as discussed before.
%However, extending \sys to such settings na\"{\i}vely, 
In this setting, as the number of containers
grows, most \sys cables are likely to be between containers.
Therefore, inter-container cables in turn require expensive optical connectors, and \sys can result in excessive cabling costs compared to a fat-tree.

%Applying \sys in this setting
%can result in excessive cabling costs compared to a fat-tree since, as the number of containers
%grows, almost all \sys cables are likely to be between containers.
%Inter-container cables require expensive optical connectors.
%Thus, for these scales, our comparisons with the fat-tree
%might be considered unfair in absence of accounting for cable lengths.

%In this section we argue that \sys can be adapted to use the same cabling cost as a fat-tree for massive data centers, while still achieving higher capacity and accomodating a higher number of servers.

However, we argue that \sys can be adapted to wire massive data centers with lower cabling cost than a fat-tree, while 
still achieving higher capacity and accommodating a larger number of servers.
For cabling the fat-tree in this setting, we apply the layout optimization suggested 
in~\cite{fattree}, \textit{i.e.,} make each fat-tree `pod' a container, and divide 
the core-switches among these pods equally. With this physical structure, we can 
calculate the number of intra-container cables (from here on referred to as
`local') and inter-container cables (`global') used by the fat-tree. We then
build a \sys network placing the same number of switches as a fat-tree pod
in a container, and using the same number of containers. \newx{The resulting \sys 
network can be seen} as a $2$-layered random graph---a random graph within each container, and
a random graph between containers. We vary the number of local and global connections 
to see how this affects performance in relation to the unrestricted \sys network. 

Note that with the same switching equipment as the fat-tree, 
\sys networks would be overprovisioned if we used the same numbers of servers.
\newest{To make sure that any loss of throughput due to our cable-optimization is clearly visible,}
we add a larger number of servers per switch to make Jellyfish oversubscribed.

\begin{figure}
\centering
\includegraphics[width=3.1in]{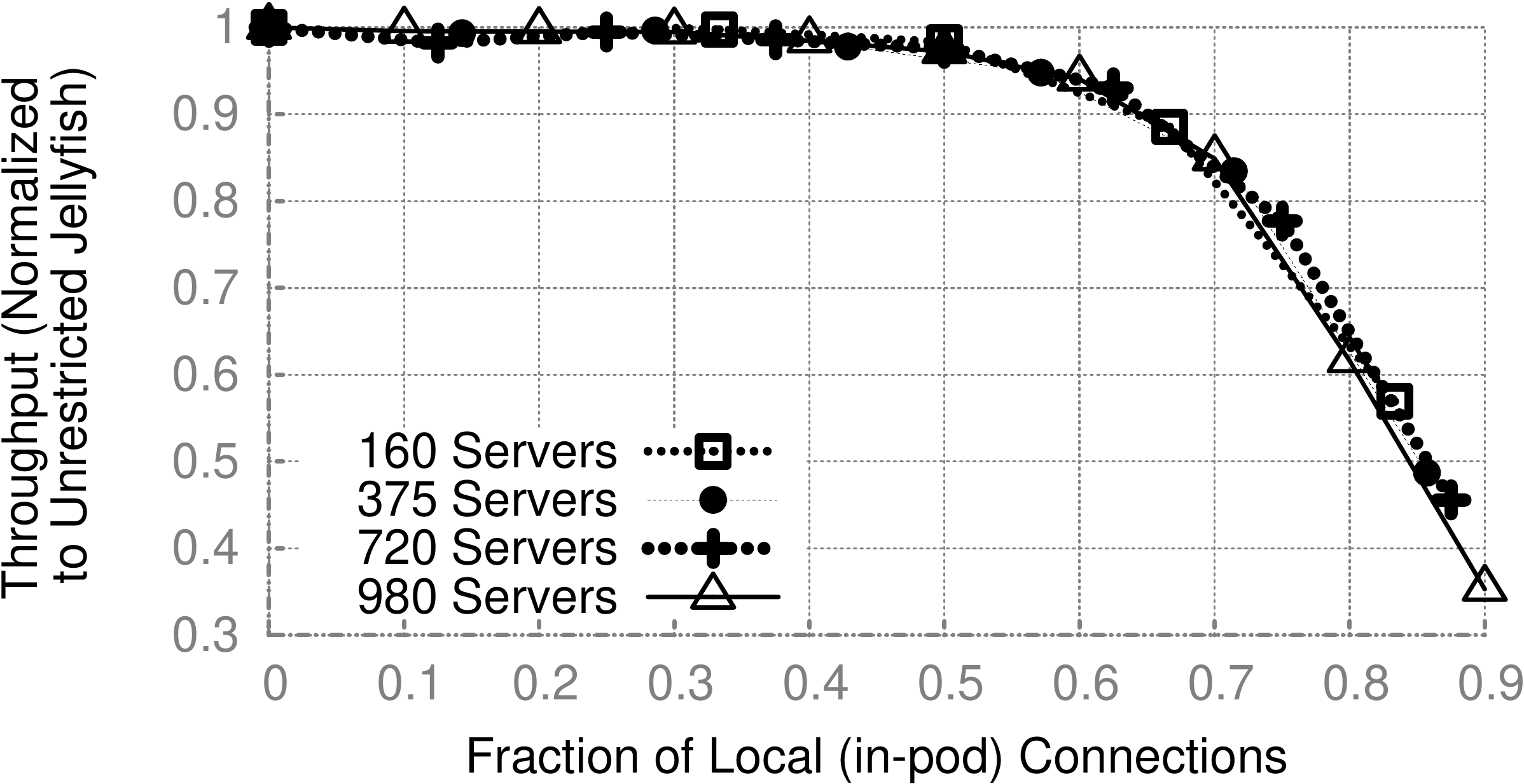}
\vspace{-8pt}
\caption{\small \em Localization of \sys random links is a promising
approach to tackle cabling for massive scale data centers: As
links are restricted to be more and more local, the network capacity
decreases (as expected). However, when $50\%$ of the random links 
for each switch are constrained to remain inside the pod, 
there is $<$$3\%$ loss of throughput.}
\label{fig:pod_plot}
\vspace{-12pt}
\end{figure}
\newest{
Fig.~\ref{fig:pod_plot} plots the capacity (average server throughput) achieved for 
$4$ sizes\footnote{These are very far from
massive scale, but these simulations are directed towards observing general trends. Much larger simulations
are beyond our simulator's capabilities.} of $2$-layer \sys, as we vary the number of local and global connections, 
while keeping the total number of connections constant for a topology. Throughput 
is normalized to the corresponding unrestricted \sys. The throughput drops by $<$$6\%$ 
when $60\%$ of the network connections per switch are `localized'. The percentage of 
local links for the equivalent fat-tree is $53.6\%$. Thus, \sys can achieve a higher
degree of localization, while still having a higher capacity network; recall 
that \sys is $27\%$ more efficient than the fat-tree at the largest scale (\S\ref{sec:efficiency}).
The effect of cable localization on throughput was similar across
the sizes we tested. For the fat-tree, the fraction of local links (conveniently given by $0.5(1 + 1/k)$ 
for a fat-tree built with $k$-port switches) \emph{decreases} marginally with size.}

\paragraphb{Complexity:} Building \sys over switches
distributed uniformly across containers will, with high probability, result in
cable assemblies between every pair of containers. A $100$,$000$
server data center can be built with $\sim$$40$ containers. Even if
\emph{all} ports (except those attached to servers) from each switch
in each container were connected to other containers, we could 
aggregate cables between each container-pair leaving us with 
$\sim$$800$ such cable assemblies, each with fewer than $200$ cables. With the 
external diameter of a $10$GBASE-SR cable being only $245um$, each such
assembly could be packed within a pipe of radius $<$$1cm$. Of course, with
higher over-subscription at the inter-container layer, these numbers could
decrease substantially.

\paragraphb{Cabling under expansion:} In massive-scale data centers,
expansion can occur through addition of new containers, or expansion
of containers (if permissible). 
%%% - not clear below
%The random connections for each layer are
%added independently by the standard \sys procedure. 
%%% - below does not seem to add much
Laying out spare cables together with the aggregates between containers 
is helpful in scenarios where a container is expanded. When a new 
container is added, new cable aggregates must be laid out to every other container.
Patch panels can again make this process easier by exposing the 
ports that should be connected to the other containers.
\section{Conclusion}
\label{sec:conclusion}
We argue that random graphs are a highly flexible architecture for
data center networks. They represent a novel approach to the significant
problems of incremental and heterogeneous expansion, while enabling
high capacity, short paths, and resilience to failures and
miswirings.

\cut{While the issues of routing and multipath 
congestion control demand more analysis, the direction appears
promising. \fixme{How specifically do they demand more analysis?  We did show they work pretty well...  Also, a big thing we should mention here is other traffic patterns.  Finally I have commented out the acknowledgements for this submission.}
}

We thank Chandra Chekuri, Indranil Gupta, Gianluca Iannaccone, Steven Lumetta, 
Sylvia Ratnasamy, Marc Snir, and the anonymous reviewers for 
helpful comments; Andy Curtis for providing code for bisection bandwidth
calculation, and sharing LEGUP topologies for comparison; and the MPTCP
authors for sharing their simulator. This work was supported in part by 
National Science Foundation grant CNS 10-40396.
 
\cut{In terms of multipath routing between top-of-rack switches, does traffic splitting 
in a flow-level granularity gives desirable throughput? 
If not, how to split packets into multipaths without triggering packet reordering problem? 
How to adapt paths in react to failures and congestion?

(section need to be reformed a bit) 
We need a way to evaluate how good a random graph. 
For expansion, we need some way to know how many switches we need to achieve a desirable properties.
It is NP-hard to derive the bisection bandwidth of a random regular graph. 
Simulation tells us how good it is, but it is time consuming... 

We have not consider any congestion control issues to the \sys. 
Wiring could be tricky in random graph... We could propose a wiring scheme that minimize the wiring cost?
While it is likely \sys provides better fault tolerance, we haven't really evaluate this.
More work on management, fault diagnosis?
}

{\bibliographystyle{abbrv}
\setlength{\bibsep}{0pt}
\footnotesize
{
%\scriptsize{
\bibliography{paper}
%}
}
}
\end{document}